\documentclass{IEEEtran}
\usepackage{amsmath,amsfonts,amssymb,eucal,graphicx}
\usepackage{wrapfig}
\usepackage{exscale}
\usepackage[usenames,dvipsnames]{color}
\usepackage{soul}
\usepackage{cite}
\usepackage{graphicx}
\usepackage{caption}
\usepackage{subcaption}
\usepackage{epstopdf}

\usepackage[export]{adjustbox}
\usepackage{tcolorbox}

\usepackage{comment}

\usepackage[capitalise]{cleveref}

\usepackage{tikz}
\usetikzlibrary{decorations.pathreplacing,calligraphy}
\usepackage[ruled,linesnumbered]{algorithm2e}
\usepackage{algorithmic}

\usepackage{float} 		
\usepackage{enumitem} 	

\usepackage{multirow}

\usepackage{wrapfig}

\usepackage{comment}

\renewcommand{\j}{\mathrm{j}}

\def\BibTeX{{\rm B\kern-.05em{\sc i\kern-.025em b}\kern-.08em
    T\kern-.1667em\lower.7ex\hbox{E}\kern-.125emX}}
\begin{document}

\title{Ray-Optics Simulations of Outdoor-to-Indoor Multipath Channels at 4 and 14 GHz}
\author{Pasi Koivum{\"a}ki, Aki Karttunen, and Katsuyuki Haneda
\thanks{P. Koivum{\"a}ki and K. Haneda are with Aalto University, Department of Electronics and Nanoengineering, 02150 Espoo, Finland. e-mail: pasi.koivumaki@aalto.fi}
\thanks{Aki Karttunen is with Tampere University, Faculty of Information Technology and Communication Sciences.}
}

\maketitle

\begin{abstract}
Radio wave propagation simulations based on the ray-optical approximation have been widely adopted in coverage analysis for a range of situations, including the outdoor-to-indoor scenario. This work presents O2I ray-tracing simulations utilizing a complete office building floor plan in the form of a laser-scanned point cloud. The simulated radio channels are compared to their measured counterparts at $4$ and $14$~GHz in terms of path loss and delay and angular spreads. Validation of channel simulations for the O2I case is rare, and so far non-existent for above-$6$~GHz bands. This work reveals the importance of a floor plan model in accurately simulating the channel; it is confirmed that path loss can be replicated with a simple interior path loss model in place of a detailed building interior model, but neglecting to model the interior results in high delay and angular spread errors. By modeling the interior, the ray-tracing simulations achieve relative mean error of under $10\%$ for delay and angular spreads. Finally, effects of multi-layer insulating window on propagation simulations are reported. Noticeable variation of the penetration loss on a small change of the incident angle of a propagation path causes large changes in estimated coverage. 
\end{abstract}

\begin{IEEEkeywords}
Point cloud, ray-tracing (RT), outdoor-to-indoor (O2I), propagation, penetration loss.
\end{IEEEkeywords}

\section{Introduction}




\IEEEPARstart{P}{roviding} wireless service of sufficient quality to indoor users is an essential goal for network operators.
Operators seek to utilize previously unused frequencies, including for example, the above-$6$~GHz new radio frequency range 2 (NR FR2)~\cite{3GPP_TR38101-2} in addition to the below-$6$~GHz legacy NR FR1~\cite{3GPP_TR38101-1} radio frequency (RF). In the legacy NR FR1, most indoor users are served by outdoor cellular infrastructure. The same service coverage becomes much more challenging in the FR2, given the higher penetration losses through e.g., building walls, experienced by radio signals. Additionally, increasing demand for energy efficiency~\cite{EUDirective} has resulted in better insulation of buildings achieved by e.g. multi-layered windows and insulating films. This has generated interest in studying indoor coverage for energy efficient smart cities of the future~\cite{LuxTurrim5G}.

To this effect, there is a continued interest in outdoor-to-indoor (O2I) channel measurements~\cite{Bas19_TWC,Karttunen_EuCAP19,Oh17_IETC,Tran16_ISAP,Imai16_EuCAP,Larsson_EuCAP14,Fuschini15_TAP}. The most commonly reported quantity is effect on signal strength inside while being serviced from outside~\cite{Imai16_EuCAP,Larsson_EuCAP14,Fuschini15_TAP}, but many studies also report large scale parameters (LSPs) of multipath channels such as delay and angular statistics~\cite{Bas19_TWC,Karttunen_EuCAP19,Tran16_ISAP,Oh17_IETC}. 

Given the difficulty involved in conducting large-scale measurement campaigns, measurement-calibrated {\it site-specific} simulations are an interesting alternative for coverage estimation. Most published results of wave propagation simulations showcase either wholly outdoor or indoor simulations instead of the O2I case, given that obtaining a complete three-dimensional (3D) model of a building can be more difficult than using exteriors obtainable from e.g. public databases. A method that has attracted recent interest is a laser-scanned point cloud of the environment used in ray-tracing~\cite{Koivumaki_TAP21,Koivumaki18_PIMRC,Jarvelainen14_RS,Jarvelainen16_TAP,Jarvelainen15_AWPL,Pang_AWPL21}. Laser-scanning can be utilized to obtain a complete model of the building and its floor plan. A number of simulation approaches have been published for O2I scenarios, e.g.~\cite{Roche10_EURASIP,Wang00_TAP,Mellios14_URSI,Degli-Esposti17_Access,DegliEsposti17_AWPL,Jimenez_CONCAPAN17}. In~\cite{Roche10_EURASIP,Wang00_TAP} ray-based propagation was combined with finite difference methods using floor plan of the building. In~\cite{Mellios14_URSI,Degli-Esposti17_Access} a path loss model was applied to indoor propagation without a model of the building interior. In~\cite{DegliEsposti17_AWPL} a ``virtual floor plan'' was generated to approximate building interior effects on propagation, while~\cite{Jimenez_CONCAPAN17} utilized a commercial ray-tracing tool with complete floor plan of the building. To the authors' best knowledge, O2I propagation simulations have so far only been compared to measurements in terms of path loss, and only for the below-$6$~GHz band by e.g.~\cite{Roche10_EURASIP,Wang00_TAP,Degli-Esposti17_Access,DegliEsposti17_AWPL,Jimenez_CONCAPAN17}. Similarly, while many approaches to O2I simulations have been published, the effects of the building interior model on LSP accuracy have not been studied. Effects of insulating structures of e.g. windows on propagation simulations and estimated coverage due to penetration loss angular selectivity is a similarly unaddressed question.

To these open questions, the novel contributions of this work are as follows:
	
	\begin{enumerate}
		
        \item Results of point cloud ray-tracing utilizing a 3D model of the building interior are presented at two frequency bands, $4$ and $14$~GHz. The frequency bands were chosen as part of LuxTurrim5G~\cite{LuxTurrim5G} to study O2I coverage at below and above-$6$~GHz bands. By comparing to measurements, path loss error is found to be in line with earlier publications reporting O2I channel simulations. Relative error of less than $10\%$ is achieved for delay and angular spreads at both bands, a result so far unaccomplished for the O2I channel.

        \item Effects of modeling the building interior on simulated channel LSPs are studied. It is shown that while path loss can be replicated with reasonable accuracy without having knowledge of the building interior, a floor plan of the building is required for accurate delay and angular spreads.

        \item Effects of a special multi-layered insulating window on the simulated channel are elaborated. It is shown that small changes of the incident angles of propagation paths cause significant changes in channel LSPs and estimated coverage due to penetration loss angular selectivity of the multi-layered windows.


  
		
	\end{enumerate}
	
The rest of the paper is organized as follows. Section~\ref{sec:O2I} describes the O2I site and its laser-scanned point cloud where spatio-temporal channel measurements were performed for validating the ray-tracing results. Section~\ref{sec:raytracing} introduces the point cloud based ray-tracing methods. Section~\ref{sec:results} presents comparisons between measured and simulated O2I radio channels. The paper is concluded in Section~\ref{sec:conclusion}.

\section{Outdoor-to-Indoor Propagation Environment}\label{sec:O2I}

This Section describes the laser-scanned point cloud model utilized in ray-tracing and the measured channel data at the same site, which were used as ground truth to optimize and validate ray-tracing results.

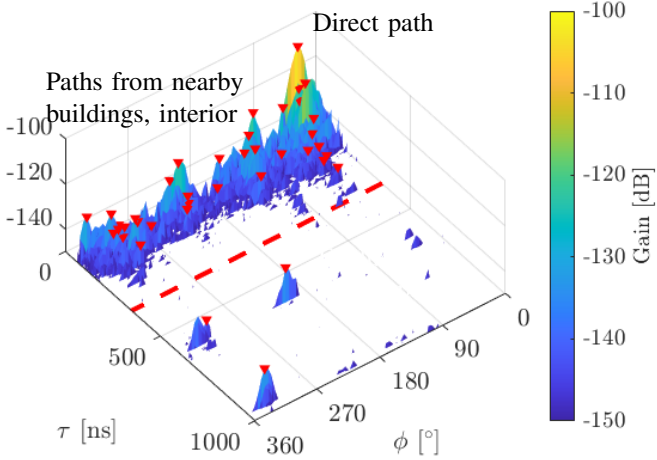
\begin{figure}[t]
    \centering
    \begin{tikzpicture}
    \node[anchor=south west,inner sep=0] (image) at (0,0) {\includegraphics[scale=0.6]{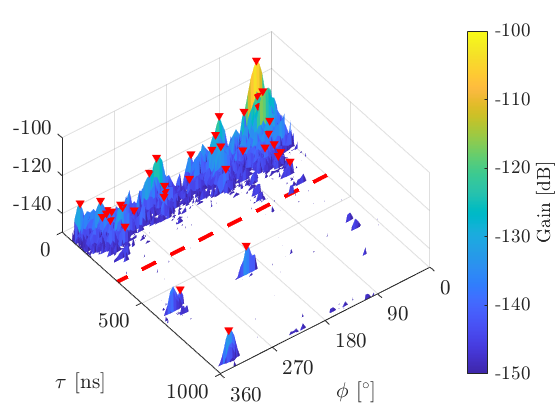}};
    \begin{scope}[x={(image.south east)},y={(image.north west)}]
        
        \node at (0.57,0.9) {Direct path};
        
        \node[rotate=0,text width=3cm] at (0.25,0.75) {Paths from nearby buildings, interior};
        
        \node[rotate=0,text width=2cm, color=white] at (0.6,0.4) {Excluded distant paths};
        
    \end{scope}
    \end{tikzpicture}
    \caption{PADP obtained at $14.25$~GHz for link Tx$2$Rx$1$. Distant paths and a limit of $350$~ns to exclude them is shown with a dashed red line.}
    \label{fig:PADP}
\end{figure}

\subsection{Channel Sounding Campaign}

Measured channels are used as a ground truth for ray-tracing. The O2I measurement campaign has been the subject of the authors' previous publications~\cite{Karttunen_EuCAP19,Koivumaki_EUCAP22}, where a more detailed description of the measurement set-up, methodology and site is provided. A total of two transmit (Tx) antenna locations and $69$ receive (Rx) antenna locations were measured at center frequencies of $4.65$ and $14.25$~GHz. The Tx locations were outside the office building and Rx locations were inside the second floor of an office building, distributed across three different rooms. The Tx antenna was elevated using a personnel lift to be on the same level with the Rx antenna. Both measurements used the same bandwidth of $500$~MHz. Directionally-revolved channel impulse responses were obtained by mechanically rotating a horn antenna on the Rx side~\cite{Koivumaki_EUCAP22}.

Figure~\ref{fig:PADP} shows an exemplary Power Angular Delay Profile (PADP) obtained from one of the links. Note that weakest gain of the PADP is limited to -150 dB, a noise threshold determined from the PADP. This is done to highlight the excluded distant paths. Signals exist below this threshold, but they are not considered meaningful to represent. For all following analysis, the studied delay range is limited to up to $\tau=350$~ns, illustrated with the dashed red line. This is to compensate for the effect of distant buildings which sometimes contribute strong propagation paths~\cite{Koivumaki_EUCAP22}. These buildings are not represented in the point cloud model used in ray-tracing, and hence measured paths from them were omitted for comparison. A propagation path is defined as a distinct local maxima in the measured PADP. A search over the PADP~\cite{Jarvelainen16_TAP,Haneda15_TAP} derived a set of discrete propagation paths to obtain comparable results to the ray-tracing simulations.

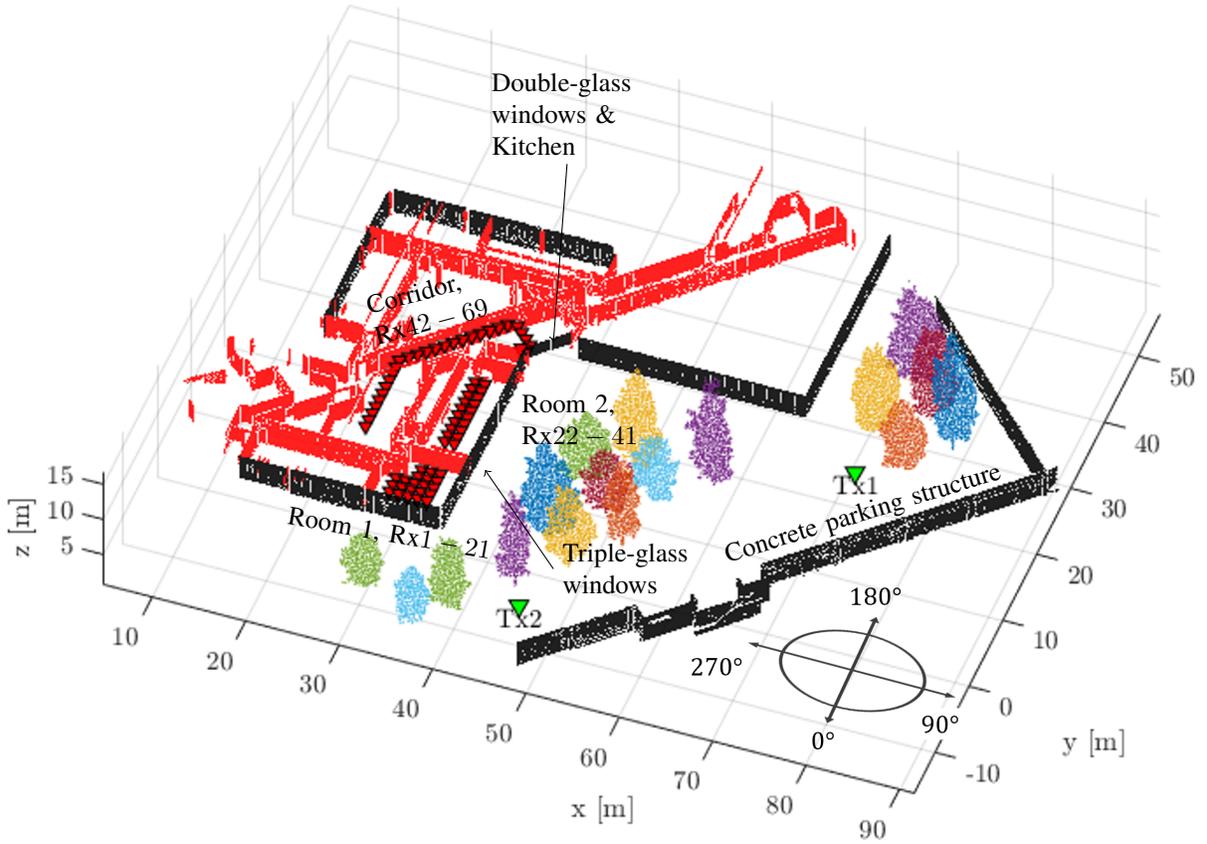
\begin{figure*}
    \centering
    \begin{tikzpicture}
    \node[anchor=south west,inner sep=0] (image) at (0,0) {\includegraphics[width=\linewidth]{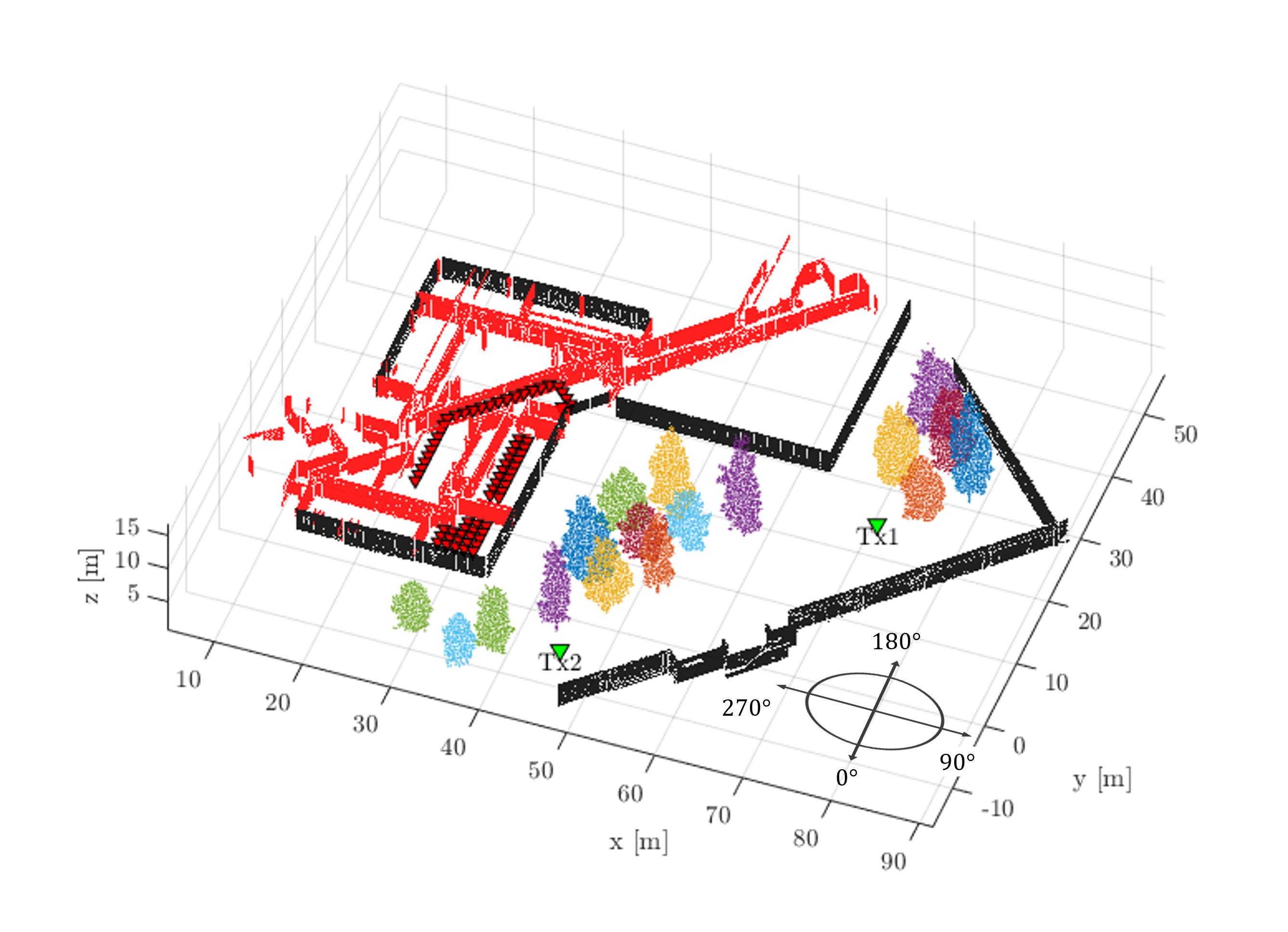}};
    \begin{scope}[x={(image.south east)},y={(image.north west)}]
        
        \node[rotate=-10] at (0.34,0.375) {Room 1, Rx$1-21$};
        
        \node[anchor=west,text width=2cm] at (0.43,0.49) {Room 2, Rx$22-41$};
        
        \node[rotate=15,text width=2cm] at (0.38,0.61) {Corridor, Rx$42-69$};

        \node[anchor=west,text width=2cm] at (0.46,0.34) {Triple-glass windows};
        \draw [->] (0.46,0.34) -- (0.41,0.44);

        \node[anchor=south,text width=2cm] at (0.47,0.75) {Double-glass windows \& Kitchen};
        \draw [->] (0.47,0.75) -- (0.46,0.57);

        \node[anchor=west,rotate=17] at (0.58,0.35) {Concrete parking structure};
    \end{scope}
    \end{tikzpicture}
    \caption{Point cloud ray-tracing model extracted from a laser-scanned point cloud. Exterior walls are shown in black, interior walls are shown in red. Trees outside the office building are shown in various colors. Reference directions of the measurement campaign are shown in degrees.}
    \label{fig:ray_tracing_model}
\end{figure*}

\subsection{Point Cloud Acquisition and Processing} \label{sec:processing}

The point clouds are captured with a Z+F IMAGER\textregistered5006h 3D laser-scanner~\cite{scanner}. The device uses movable mirrors to steer a laser beam in different directions to detect distances to reflective surfaces. A number of locations outside the building and inside on the $2^{\rm nd}$ floor are scanned and combined into a complete model of the environment. Resolution of the point cloud used in this work is approximately $10$~cm. To obtain a point cloud appropriate for ray-tracing simulations, the following steps were performed.

\begin{enumerate}
    \item Point clouds obtained outside and inside the office building were aligned and merged into one complete point cloud using common reference points.
    
    \item Vertical interior walls of the second floor and the exterior walls on the level of the Tx-Rx links are extracted from the laser-scanned point cloud by detecting large flat sections~\cite{Koivumaki_TAP21}. They are shown in Fig.~\ref{fig:ray_tracing_model} in red and black, respectively, the black wall opposite to the office building being a parking structure. Ceilings and floors of the $2^{\rm nd}$ floor are removed along with the ground outside the building to reduce the size of the point cloud.
    
    \item Individual trees and their canopies are extracted from the laser-scanned point cloud manually. They are shown in Fig.~\ref{fig:ray_tracing_model} in various colors.
\end{enumerate}

The complete point cloud model used in ray-tracing is shown in Fig.~\ref{fig:ray_tracing_model}. All $69$ measured Rx locations across three different rooms are shown with red triangles. Room~$1$ is a square corner room with triple-glass windows facing the outside housing Rx locations $1$-$21$. Room~$2$ is a rectangular room with triple-glass housing Rx locations $22$-$41$. The third area consists of a kitchen with a double-glass window facing the outside and a corridor that runs behind rooms 1 and 2, housing Rx locations $42$-$69$. The exterior walls with triple- and double-glass windows are highlighted in Fig.~\ref{fig:ray_tracing_model}.

\section{Point Cloud Ray-Tracing}\label{sec:raytracing}

This Section describes the ray-tracing methods for determining propagation paths between the Tx and Rx. Gains of the traced paths are estimated separately as introduced in Section~\ref{sec:results}. The direct propagation path between Tx and Rx along with specular reflections are considered. Each traced path was subject to determine if it undergoes shadowing due to building walls and vegetation. 



\subsection{Direct Path}

The direct path between Tx and Rx is determined with the Tx and Rx locations illustrated in Fig.~\ref{fig:ray_tracing_model}. The Tx constitutes a starting point of the propagation path and the Rx its ending point.

\begin{figure}[t]
    \centering
    \includegraphics[scale=1]{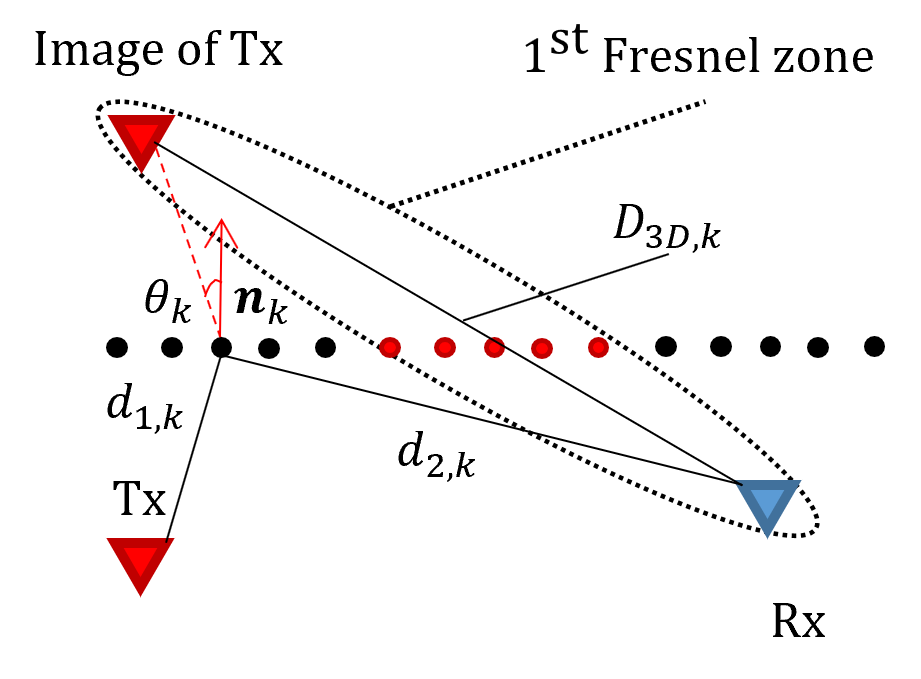}
    \caption{Detecting a single-bounce specular reflection from a point cloud. Points which satisfy Eq.~\eqref{eq:specular} are colored with red.}
    \label{fig:specular}
\end{figure}

\subsection{Specular Reflections} Specular reflection is an interaction of a plane wave with an electrically large surface where the angles of incidence and departure are equal. Our method for detecting specular reflections in a point cloud environment is based on an established technique~\cite{Virk_LAPC15,Jarvelainen16_TAP,Koivumaki_TAP21}, which utilizes the image method and the $1^{\rm st}$ Fresnel zone. Detection of single-bounce specular reflections from a section of a point cloud is illustrated in Fig.~\ref{fig:specular}. An image of the Tx is calculated for each point in the point cloud using its normal vector. To find all valid single-bounce reflected paths, it is determined if the point lies within the $1^{\rm st}$ Fresnel zone between the image Tx and Rx, i.e. that
\begin{equation}
    d_{1,k} + d_{2,k} - D_{{\rm 3D},k} \leq \frac{\lambda}{2}, \label{eq:specular}
\end{equation}
where $D_{{\rm 3D},k}$ is distance between the Rx and image of Tx corresponding to the $k^{\rm th}$ point. As shown in Fig.~\ref{fig:specular}, multiple closely located points in a single flat section of the point cloud may fulfill Eq.~\eqref{eq:specular}. To avoid seeing multiple reflections from a single surface, reflection points are grouped as described in~\cite{Jarvelainen16_TAP}. For higher order reflections, the image method is continued until the desired number of bounces is reached.

\subsection{Detection of Shadowing Events}\label{sec:detect_shadow}

\begin{figure}
\begin{subfigure}[b]{\linewidth}
\begin{center}
    \includegraphics[scale=1]{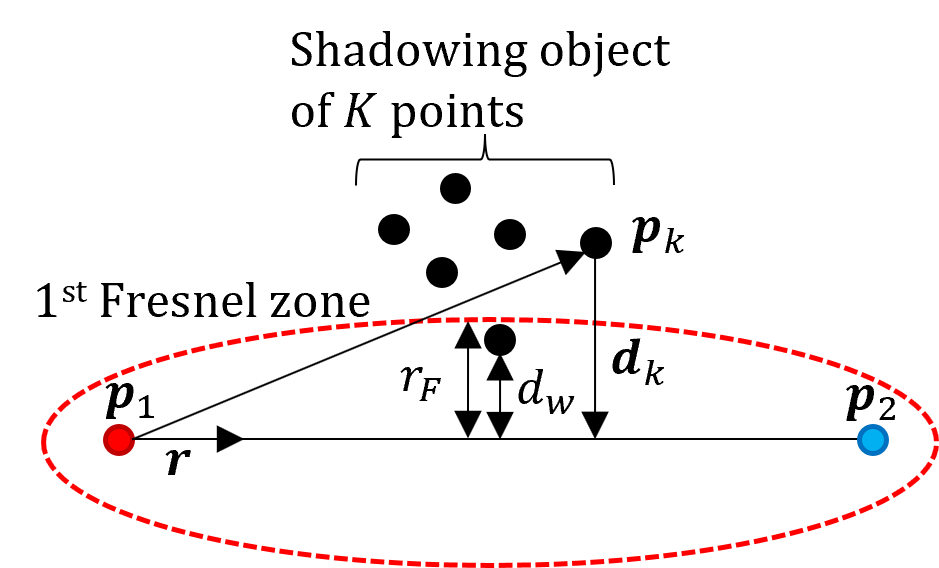}
    \caption{}
    \end{center}
    \end{subfigure}
    
    \begin{subfigure}[b]{\linewidth}
    \begin{center}
    \includegraphics[scale=1]{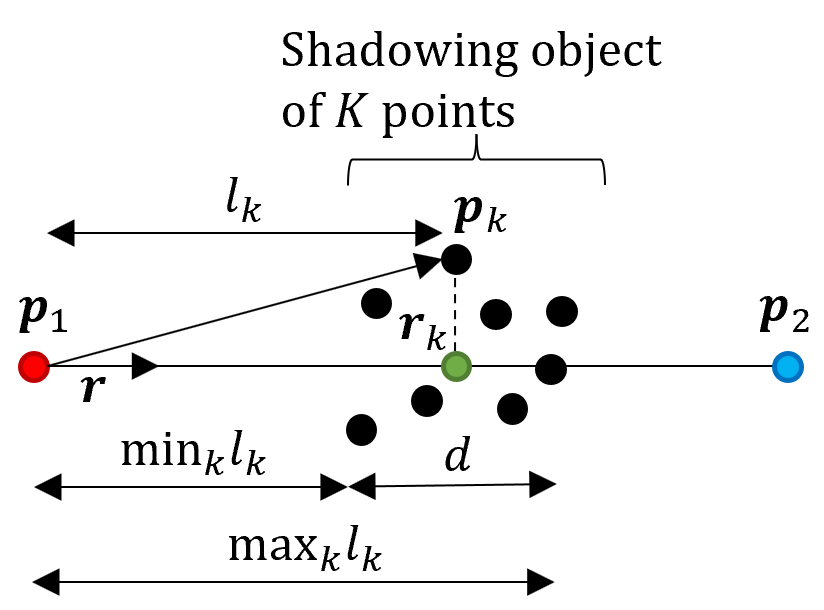}
    \caption{}
    \end{center}
    \end{subfigure}

    \caption{Calculating distance from an object to a ray (a) and propagation distance inside an object (b).}
    \label{fig:shadowing}
\end{figure}

Detecting shadowing events of a propagation path in a point cloud has been presented in \cite{Jarvelainen_letter16, Koivumaki_TAP22} to derive the line-of-sight probability. A propagation path is assumed to be shadowed by an object, consisting of $K$ points whose position vectors are $\boldsymbol{p}_k$, $1 \leq k \leq K$, if there are points inside the $1^{\rm st}$ Fresnel ellipsoid. The previous inequality in Eq.~\eqref{eq:specular} can be applied, where $d_{1,k}$ and $d_{2,k}$ are lengths of the propagation path from a starting point $\boldsymbol{p}_1$ via $k^{\rm th}$ point in the point cloud to the ending point $\boldsymbol{p}_2$. Similarly, $D_{{\rm 3D},k}$ is the distance between $\boldsymbol{p}_1$ and $\boldsymbol{p}_2$ and $\lambda$ is the wavelength. If the inequality in Eq.~\eqref{eq:specular} is satisfied by any point $1 \leq k \leq K$, a point of the object is within the $1^{\rm st}$ Fresnel zone and the propagation path is considered shadowed.

Leveraging high level-of-detail inherent to laser-scanned point clouds, two influential geometrical parameters are defined to calculate penetration losses. 

\begin{enumerate}
    \item Distance from the ray to nearby surrounding objects that may shadow the ray, $d_w$. For a single object consisting of $K$ points, 
    the distance between the object and ray is given by $d_w = \min_k | \boldsymbol{d}_k |$, where 
    \begin{equation}
    \boldsymbol{d}_k = \boldsymbol{p}_{1} - \boldsymbol{p}_k - ((\boldsymbol{p}_{1} - \boldsymbol{p}_k ) \cdot \boldsymbol{r})\boldsymbol{r}
    \end{equation}
    is a vector projecting the point $\boldsymbol{p}_k$ onto the ray, whose offset and propagation direction is given by $\boldsymbol{p}_{1}$ and $\boldsymbol{r}$, respectively; the operator $(\cdot)$ represents an inner product. The defined ray-object distance is used to obtain the penetration loss estimates by introducing a heuristic scaling factor
    \begin{equation}
        q = 1 - \frac{d_w}{r_{\rm F}},
    \end{equation}
    where $r_{\rm F}$ is radius of the $1^{\rm st}$ Fresnel zone at the object. Penetration losses from tree canopies and interior walls are scaled using $q$ to account for the changing size of the $1^{\rm st}$ Fresnel zone at different frequencies and propagation distances as detailed in Section~\ref{sec:loss}. The process is illustrated in Fig.~\ref{fig:shadowing}(a).
    
    \item Path length $d$ over which a ray undergoes penetration into an object. Projecting the point $k$ in an object onto the ray, its position vector is given by 
    \begin{equation}
    \boldsymbol{r}_k = \boldsymbol{p}_1 - ((\boldsymbol{p}_1 - \boldsymbol{p}_k ) \cdot \boldsymbol{r})\boldsymbol{r}.
    \end{equation}
    Its distance along the ray from the Tx antenna is $l_k = |\boldsymbol{p}_{1} - \boldsymbol{r}_k|$. With these definitions, the penetration length along the path is given by
    \begin{equation}
    d = \max_k l_k - \min_k l_k, \label{eq:ds}
    \end{equation}
    for points blocking the ray. Once again, $K$ points constitute a single blocking object. The path length is used to apply tree canopy losses based on a per-meter attenuation as described in Section~\ref{sec:loss}. The process is illustrated in Fig.~\ref{fig:shadowing}(b).
\end{enumerate}

\section{Comparison Between Ray-Tracing Simulations and Measurements}\label{sec:results}

Having traced rays and defined several influential geometrical parameters in the previous Section, the method to estimate gains of each traced paths is defined in this Section. Then the results of ray-tracing simulations are compared against the measured propagation channel in terms of its LSPs.

\begin{table}[t]
    \centering
    \caption{Electrical properties of materials.
    \label{tab:materials}}
    \begin{tabular}{l c c}
    \hline\hline
        Frequency band   &  $4$~GHz & $14$~GHz\\  \hline
        Concrete    & $5.31+\j0.45$ & $5.31+\j0.35$\\
        Plasterboard& $2.94+\j0.14$ & $2.94+\j0.09$\\
        Glass       & $6.27+\j0.10$ & $6.27+\j0.13$\\ 
        Metal       & $1+\j4.50\times10^8$ & $1+\j1.28\times10^8$\\ 
        
        \hline\hline
    \end{tabular}
\end{table}

\subsection{Ray-Tracing Simulation Set-Up}

\begin{figure}[t]
    \begin{center}
    \begin{subfigure}[b]{0.49\linewidth}
    \includegraphics[scale=1]{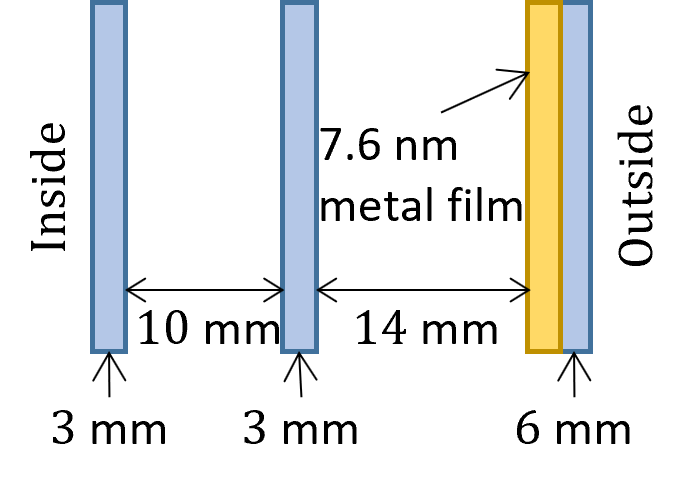}
    \caption{}
    \end{subfigure}
    \end{center}
    
    \begin{subfigure}[t]{0.49\linewidth}
    \begin{center}
    \includegraphics[scale=1]{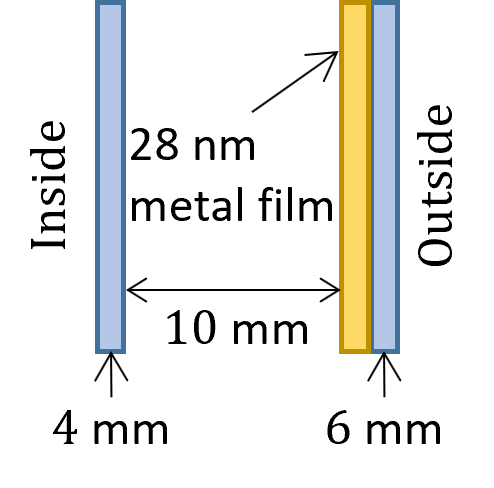}
    \end{center}
    \caption{}
    \end{subfigure}
    \begin{subfigure}[t]{0.49\linewidth}
    \begin{center}
    \includegraphics[scale=1]{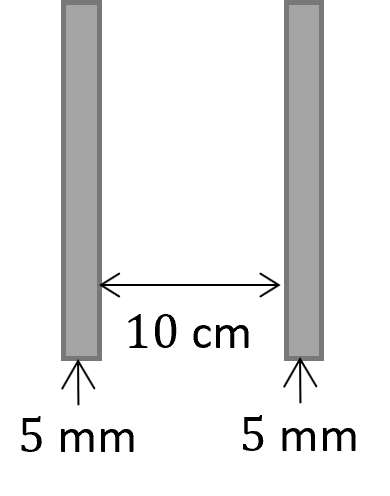}
    \end{center}
    \caption{}
    \end{subfigure}
    \caption{Triple-glass (a) and double-glass (b) windows with insulating metallic films, and double-plasterboard interior wall (c). Thickness of each layer is not to scale. \label{fig:windows}}
\end{figure}

The following assumptions are made regarding structures observed at the measurement site to assign permittivity values to different parts of the environment. Windows of the office building were noted to consist of triple- and double-glass windows as shown in Figs.~\ref{fig:windows}(a) and (b), respectively. The layered window structures were not modeled in the raw point cloud and hence manually measured and modeled by hands. A thin metallic film, likely for added insulation, was known to exist on the interior side of the outermost window. Based on laboratory measurements, effective thicknesses of the metallic films were known to be $7.6$~nm and $28$~nm for the triple- and double-glass windows, respectively. Exact structure of the insulating metal film company-proprietary and hence is not known to the authors, and thus an effective thickness is used.

The simulated penetration loss through the triple-glass window is shown in Fig.~\ref{fig:transmission_loss}. The loss through a double-plasterboard interior wall, which does not use a metallic film, is overlaid. Mean penetration losses across each frequency band are shown. The mean value is used in the ray-tracing simulations. At the $14$~GHz band the loss through a triple-glass window oscillates significantly. A change of approximately $2^\circ$ in incident angles of a plane wave can result in penetration loss difference of up to $20$~dB. The oscillation is not present at the $4$~GHz band, nor for the double-glass window which is not shown. This is a consequence of the simulated material parameters and layered structure, combined with wavelength. Significant constructive and destructive interference happens only at the 14 GHz band. Penetration loss through the interior walls has a similar level for both frequency bands.

Interior walls of the building separating office spaces were assumed to be typical plasterboard walls consisting of two layers. This structure is shown in Fig.~\ref{fig:windows}(c). Exterior walls of the building are assumed to consist only of windows for simplicity, although there are wooden window frames and some concrete supporting structures included in the facade. The parking structure located opposite to the office building is assumed to consist of concrete.

Well-accepted ITU-R recommendation P.2040~\cite{ITU-R_P2040} provides permittivity values and formulas to calculate reflection and transmission coefficients using the multi-layer slab model. Permittivity values used in ray-tracing simulations are reported in Table~\ref{tab:materials}. Roughness of reflecting surfaces is not considered to affect the calculated coefficients.

Specular reflections up to four bounces are simulated. Diffractions and diffuse scattering are not simulated. Due to being uneven surfaces, trees are assumed not to be sources of important propagation paths. Only interior and exterior walls shown in Fig.~\ref{fig:ray_tracing_model} are considered as sources of reflected paths. Ceilings and floors of the building are not considered as sources of reflections. Any scatterers, e.g. metallic piping concealed by the false ceiling of the office are not included in the model as they are not visible to the laser-scanner. While performing ray-tracing simulations, two different point cloud models are considered. The first consist of {\it only exterior walls} of the office building and nearby buildings. As interior structure of the building is assumed unknown in this case, a distance-dependent path loss model is applied to propagation inside the building~\cite{ITU-R_M2135-1}. The second model in addition includes all interior walls of the $2^{\rm nd}$ floor, referred to as {\it full floor plan} hereinafter, to test their importance in reproducing the measured propagation channel and its characteristics. For both point clouds a resolution of 10 cm between points is used to guarantee that specular reflections can be modeled from each surface as illustrated in Fig.~\ref{fig:specular}.

\subsection{Obtaining Tree Canopy Loss}\label{sec:loss}

\begin{figure}[t]
    \centering
    \includegraphics[scale=0.5]{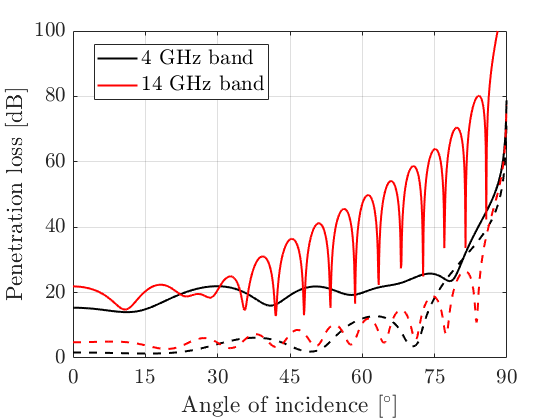}
    \caption{Simulated mean penetration loss of the triple-glass window (solid lines) and a double-plasterboard interior wall (dashed lines) across each studied band. Values obtained using configurations in Fig.~\ref{fig:windows} and permittivity values from Table~\ref{tab:materials}.}
    \label{fig:transmission_loss}
\end{figure}

To estimate propagation loss through tree canopies, the direct connection paths between Tx and Rx antennas are analyzed. Note that while well-accepted models exist for tree canopy attenuation~\cite{ITU-R_P833}, new values are estimated in this work to obtain the best result. Attenuation in vegetation is known to be a highly site-specific phenomenon. Their excess losses to the free space losses are of interests because they are attributed to penetration of the direct path through different window types, interior walls dividing office spaces and tree canopies. The tree losses are estimated by minimizing the difference between measured and simulated excess losses of all direct connection paths. Separate values are determined for tree canopy loss at the two frequency bands.

Propagation delays $\tau_{\rm d}$ and azimuth angles of arrival $\phi_{\rm d}$ of direct paths are determined geometrically from the Tx-Rx floor plan of the measurements. Note that due to the propagation environment, it is not feasible to assume that the direct path is always the strongest path. Reflected paths can sometimes be significantly stronger. Due to this, the Tx-Rx coordinates have to be used in a more detailed peak search. The reading of gains in the measured PADPs at the delay and azimuth angle serves as the direct path gain estimate. Specifically, fine estimates of the delay and azimuth angle of arrival of the direct path are identified using the measured PADP as 
\begin{eqnarray}
    (\hat{\tau_{\rm d}}, \hat{\phi_{\rm d}}) & = & \arg \max_{\substack{\tau_{\rm d}-\Delta\tau \le \tau \le \tau_{\rm d}+\Delta\tau, \\ \phi_{\rm d}-\Delta\phi \le \phi \le \phi_{\rm d}+\Delta\phi}} {\rm PADP} (\tau,\phi), \\
    \hat{G}_{\rm d} & = &  {\rm PADP} (\hat{\tau}_{\rm d}, \hat{\phi}_{\rm d}),
\end{eqnarray}
where $\hat{\cdot}$ indicates an estimate of corresponding variable and $\Delta\tau$ and $\Delta\phi$ define delay and azimuth ranges over the PADP to find a local maximum. The fine estimates are required to account for uncertainty of the Tx and Rx coordinate information which are manually obtained during measurements. Search ranges in the azimuth $\Delta \phi = 5^\circ$ and in the delay $\Delta \tau = 2$~ns are chosen, both corresponding the their respective resolutions of the channel sounding. Measured excess loss of the direct path is estimated by subtracting the free space path loss as
\begin{equation}
    L_{\rm ex} \; {\rm [dB]}= -10 \log_{10} \hat{G}_{\rm d} - 10 \log_{10} \biggr( \frac{1}{4\pi \hat{\tau}_{\rm d} f_{\rm c}} \biggr).
\end{equation}
The excess loss simulated with point cloud ray-tracing is given by the generic formula
\begin{equation}
\begin{split}
    &L_{\rm ex, sim} \; {\rm [dB]} =\\
    &\sum^{N_{\rm wdw,1}}_{i=1} L_{\rm wdw,1}(\theta_{\rm wdw,1,i}) + \sum^{N_{\rm wdw,2}}_{i=1} L_{\rm wdw,2}(\theta_{\rm wdw,2,i}) + \\
    &\sum^{N_{\rm tree}}_{i=1} L_{\rm tree} \cdot d_{{\rm tree},i} \cdot q_{{\rm tree},i} +  \sum^{N_{\rm iw}}_{i=1} L_{\rm iw}(\theta_{{\rm iw},i}) \cdot q_{{\rm iw},i},
\end{split}\label{eq:excess}
\end{equation}
where $L_{{\rm wdw},k}$, $k=1,2$ is penetration loss through the triple- and double-glass windows, respectively. Total window penetration losses through $N_{{\rm wdw},k}$ windows are calculated using~\cite{ITU-R_P2040} and the angle of incidence $\theta_{{\rm wdw},k,i}$, where the normal and grazing incidence to the window corresponds to $0^\circ$ and $90^\circ$, respectively. Penetration losses through $N_{\rm tree}$ tree canopies are calculated using the canopy loss $L_{\rm tree}$, propagation distance inside the canopy $d_{{\rm tree},i}$ and heuristic scaling factor $q_{{\rm tree},i}$ described in Section~\ref{sec:detect_shadow}. Penetration losses through $N_{\rm iw}$ interior walls are scaled similarly, where $L_{\rm iw}$ is calculated using angle of incidence $\theta_{\rm iw,i}$ and multi-layer slab model~\cite{ITU-R_P2040}. Note that in the case of no penetrations of a particular environmental feature its respective term in Eq.~\eqref{eq:excess} is zero.

\begin{figure}[t]
    \centering
    \begin{subfigure}[b]{\linewidth}
    \includegraphics[width=\linewidth]{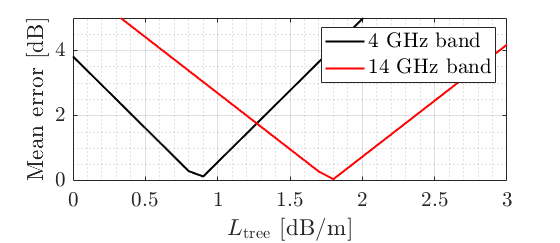}
    \caption{}
    \end{subfigure}
    \begin{subfigure}[b]{\linewidth}
    \includegraphics[width=\linewidth]{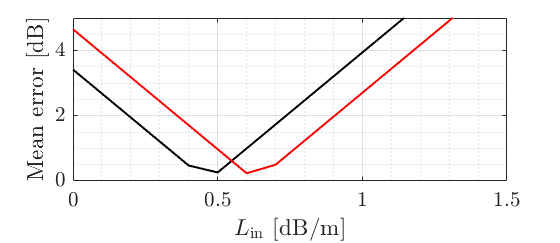}
    \caption{}
    \end{subfigure}
    \caption{Determining the (a) tree canopy loss, used in Eqs.~\eqref{eq:excess},\eqref{eq:excess_din}, and (b) interior propagation loss, used in Eq.~\eqref{eq:excess_din}, by minimizing direct path excess loss mean error.}
    \label{fig:treeloss}
\end{figure}

Tree canopy losses $L_{\rm tree}$ are obtained by minimizing the mean error between simulated and measured direct path excess losses. The effect of different $L_{\rm tree}$ on the simulated mean error is shown in Fig.~\ref{fig:treeloss}(a). The lowest errors, approximately $0.1$~dB, are achieved with canopy losses of $0.9$ and $1.8$~dB/m for the $4$ and $14$~GHz bands, respectively. The values are in line with what is recommended in~\cite{ITU-R_P833}.

\begin{figure*}
\begin{center}
    \begin{subfigure}[b]{\linewidth}
    \begin{tikzpicture}
    \node[anchor=south west,inner sep=0] (image) at (0,0) {\includegraphics[width=\linewidth]{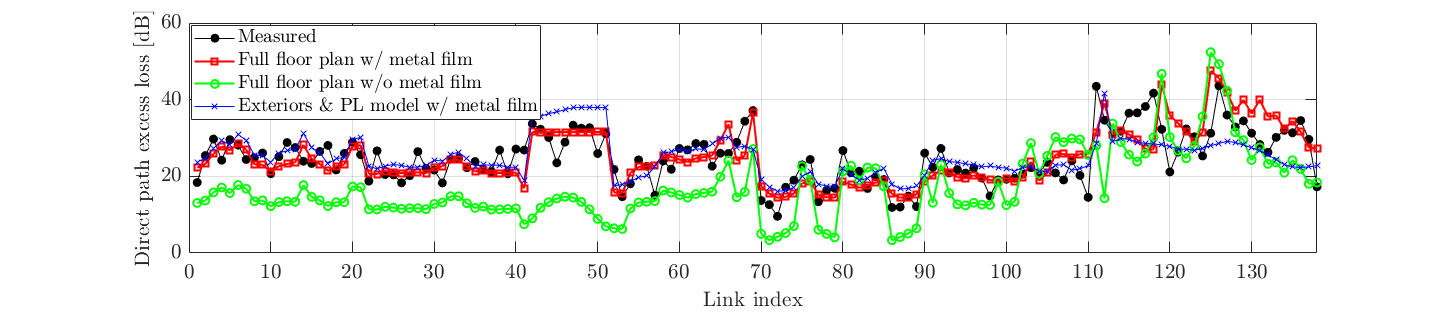}};
    \begin{scope}[x={(image.south east)},y={(image.north west)}]

        \draw [->] (0.525,0.75) -- (0.555,0.75);
        \draw [<-] (0.525,0.75) -- (0.495,0.75);
        \draw [-] (0.525,0.71) -- (0.525,0.79);

        \node[anchor=south] at (0.55,0.75) {Tx2};
        \node[anchor=south] at (0.50,0.75) {Tx1};
    \end{scope}
    \end{tikzpicture}
    \caption{}
    \end{subfigure}
    \begin{subfigure}[b]{\linewidth}
    \begin{tikzpicture}
    \node[anchor=south west,inner sep=0] (image) at (0,0) {\includegraphics[width=\linewidth]{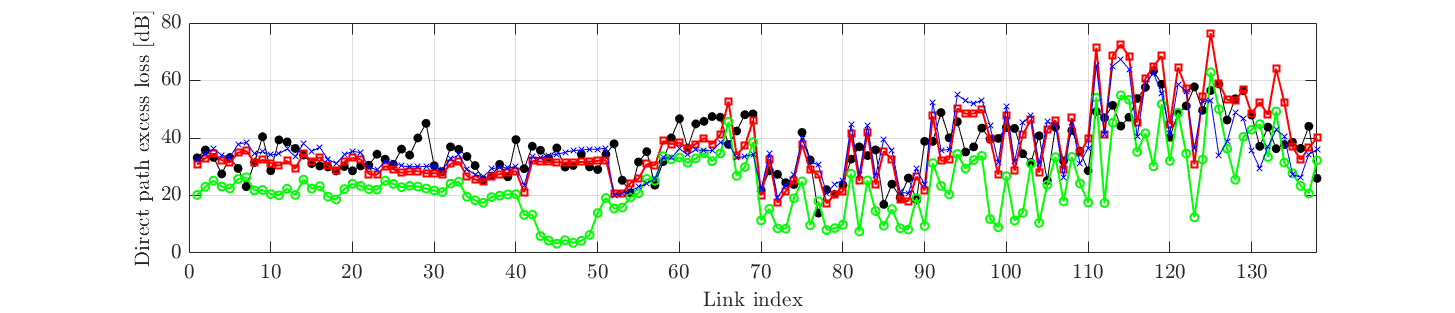}};
    \begin{scope}[x={(image.south east)},y={(image.north west)}]

        \draw [->] (0.525,0.75) -- (0.555,0.75);
        \draw [<-] (0.525,0.75) -- (0.495,0.75);
        \draw [-] (0.525,0.71) -- (0.525,0.79);

        \node[anchor=south] at (0.55,0.75) {Tx2};
        \node[anchor=south] at (0.50,0.75) {Tx1};

        \node[anchor=south] at (0.26,0.7) {Concrete pillar};
        \draw[->] (0.26,0.7)--(0.276,0.6);
    \end{scope}
    \end{tikzpicture}
    \caption{}
    \end{subfigure}
    \end{center}
    \caption{Measured and simulated excess loss of the direct path at $4$~GHz (a) and $14$~GHz (b) bands. Simulated excess loss calculated using obtained tree canopy loss and window metal film thicknesses.}
    \label{fig:direct_ray_excess}
\end{figure*}
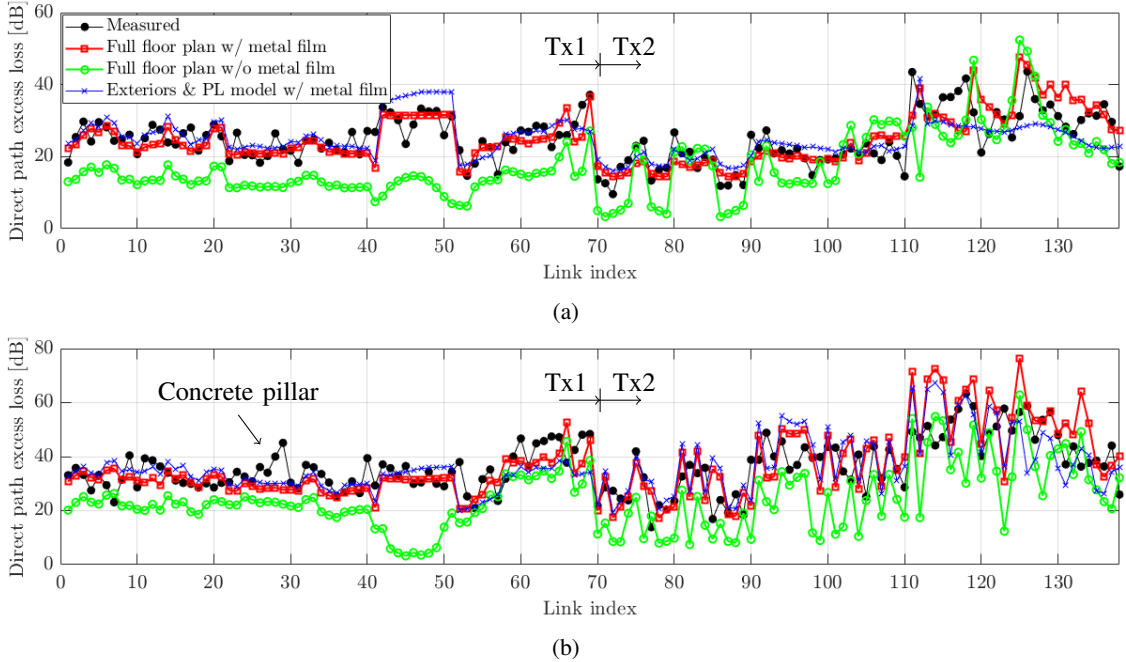

The measured and full floor plan simulated excess losses of direct paths are shown in Fig.~\ref{fig:direct_ray_excess}. The excess losses are shown with and without the insulating metal films. The former follows trends of the measured direct path excess loss, while the latter fail to reproduce the trend in measurements. Note that sometimes no metal film results in higher losses than with metal film. This is the effect of the simulated layered materials. It is clear that inclusion of the metal film was critical in accurately reproducing measurements.

The measured and simulated excess losses fluctuate strongly for links that belong to Tx2, i.e. $70$-$138$. This because angles of incidence through the triple-glass windows are greater than $45^\circ$, evident from Fig.~\ref{fig:ray_tracing_model}. The oscillating penetration loss through the window is seen in Fig.~\ref{fig:transmission_loss}, indicating high selectivity based on angle of incidence. This can be used to explain the difference to measurements for the $14$~GHz band, as there is some uncertainty between exact Tx and Rx antenna locations. Even small difference between simulated and actual angle can result in a large change in penetration loss. Overall difference can be attributed to the trees being modeled as homogeneous. Some environmental details are missing from the simulation model, e.g. wooden supporting structures of the windows and structures inside the building. The effect of a concrete support pillar in the facade not included in the model is pointed out for the $14$~GHz band.

\subsection{Obtaining Interior Propagation Loss}

To estimate propagation loss inside the building with a per-meter loss, instead of considering wave interaction with interior walls, the direct connection paths between Tx and Rx antennas are analyzed. The procedure is the same as for obtaining the tree canopy loss. In this case, the excess loss simulated with point cloud ray-tracing is given by the generic formula
\begin{equation}
\begin{split}
    &L_{\rm ex, sim} \; {\rm [dB]} =\\
    &\sum^{N_{\rm wdw,1}}_{i=1} L_{\rm wdw,1}(\theta_{\rm wdw,1,i}) + \sum^{N_{\rm wdw,2}}_{i=1} L_{\rm wdw,2}(\theta_{\rm wdw,2,i}) + \\
    &\sum^{N_{\rm tree}}_{i=1} L_{\rm tree} \cdot d_{{\rm tree},i} \cdot q_{{\rm tree},i} + L_{\rm in} \cdot d_{\rm in},
\end{split}\label{eq:excess_din}
\end{equation}
where the new variables $L_{\rm in}$, a dB/m interior propagation loss, and $d_{\rm in}$, propagation distance inside the building, replace the losses from interior walls. Tree canopy loss determined in Section~\ref{sec:loss} is used.

The effect of different $L_{\rm in}$ on the simulated mean error is shown in Fig.~\ref{fig:treeloss}(b). The lowest errors, approximately $0.25$~dB, are achieved with interior losses of $0.5$ and $0.6$~dB/m for the $4$ and $14$~GHz bands, respectively. The values are in line with the frequency-independent $0.5$ dB/m recommended in~\cite{ITU-R_M2135-1}, although it is for a Manhattan grid layout.

The simulated direct path excess losses obtained with building exteriors and the interior distance-dependent path loss model are shown in Fig.~\ref{fig:direct_ray_excess}. The simulated values follow the measurements well, except for link indices $110$-$138$ at $4$~GHz. The Rx is deep inside the building here, and the distance-dependent path loss does not reproduce these particular cases well. 

\subsection{Comparison Metrics}

Having obtained the tree penetration and interior propagation losses using direct paths, path loss of the channel is calculated to study efficacy of the ray-tracing simulation. The path loss of a link is derived by summing gains of all traced paths up for the link and taking a base-10 logarithm of it. Angular and delay spreads of the channels are calculated according to~\cite{Molisch11_book} as well to evaluate efficacy of the ray-tracing simulation in terms of multipath richness in the angular and delay domains. Visual comparisons of the measured and simulated power angular profile (PAP) and power delay profile (PDP) are also given. 
A dynamic range of $20$~dB from the strongest propagation path is used for both measured and simulated channels when calculating LSPs. 

\begin{figure}[t]
    \centering
    \begin{tikzpicture}
    \node[anchor=south west,inner sep=0] (image) at (0,0) {\includegraphics[width=\linewidth]{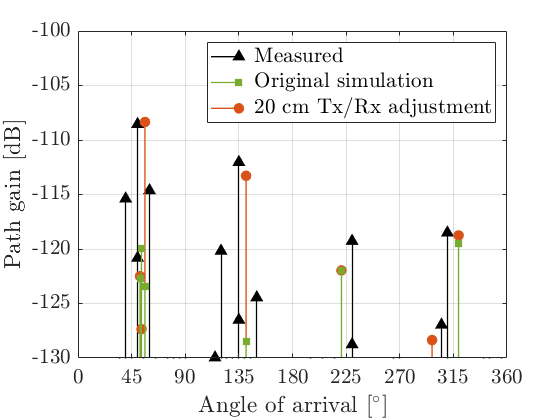}};
    \begin{scope}[x={(image.south east)},y={(image.north west)}]
    
        \node[anchor=south] at (0.26,0.75) {Path~$1$};
        
        \node[anchor=south] at (0.43,0.64) {Path~$2$};
        
        \node[anchor=south] at (0.63,0.5) {Path~$3$};
        \node[anchor=south] at (0.8,0.5) {Path~$4$};
        

    \end{scope}
    \end{tikzpicture}
    \caption{Effect of a $20$~cm adjustment of simulated Tx and Rx12 locations on $14$~GHz channel PAP. Paths of interest are indicated.}
    \label{fig:adjusted}
\end{figure}
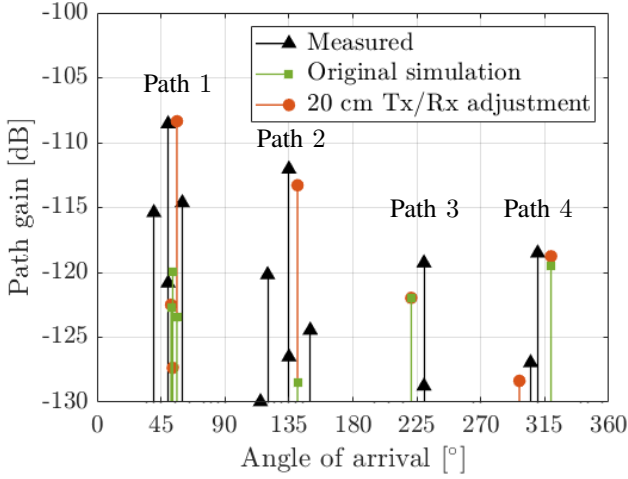

\begin{figure}[t]
    \includegraphics[width=\linewidth]{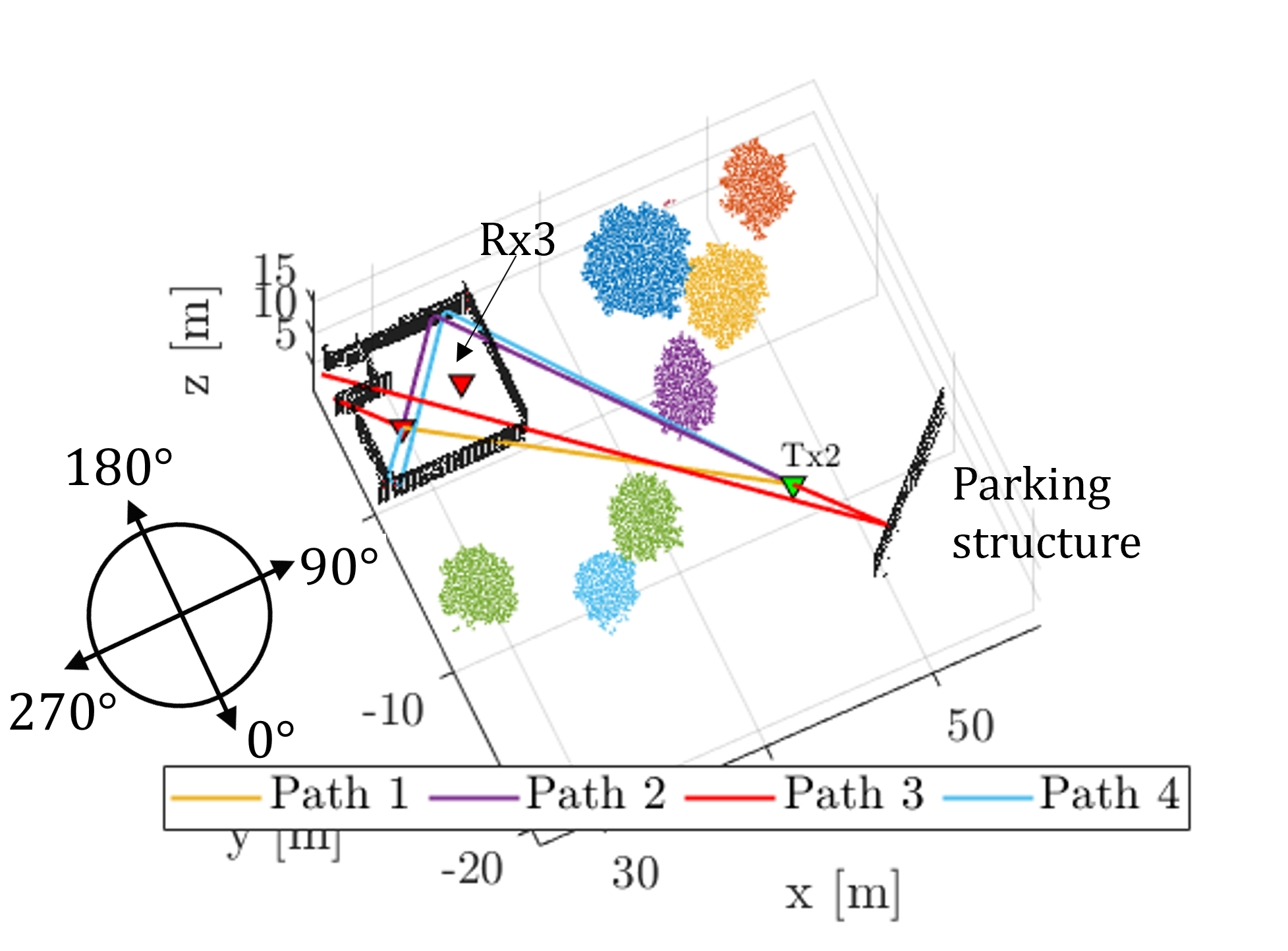}
    \caption{A closer view of Room~$1$ and Tx$2$. The $4$ paths from Fig.~\ref{fig:adjusted} are drawn between Tx$2$ and Rx$12$.}
    \label{fig:zoomed_room1}
\end{figure}

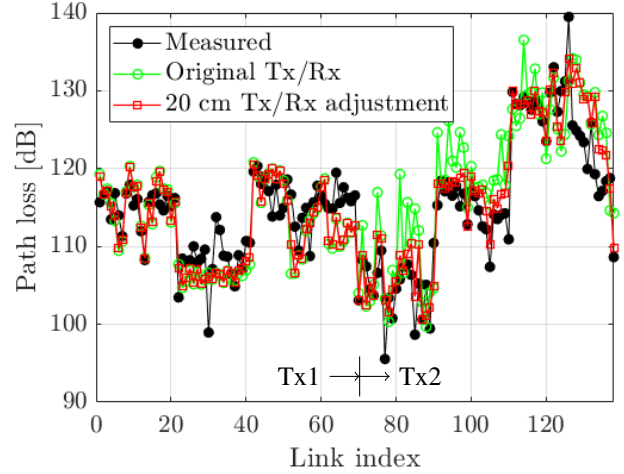
\begin{figure}[t]
\begin{tikzpicture}
    \node[anchor=south west,inner sep=0] (image) at (0,0) {\includegraphics[width=\linewidth]{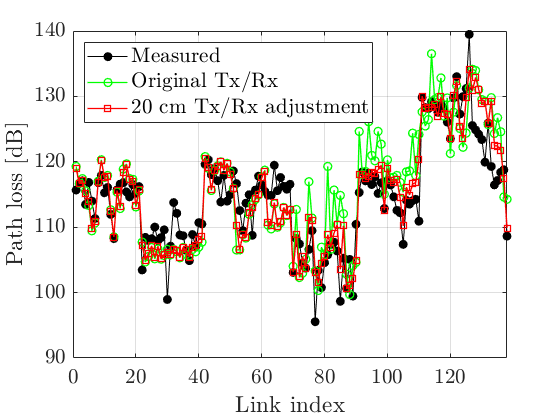}};
    \begin{scope}[x={(image.south east)},y={(image.north west)}]

        \draw [->] (0.525,0.20) -- (0.57,0.20);
        \draw [<-] (0.525,0.20) -- (0.48,0.20);
        \draw [-] (0.525,0.16) -- (0.525,0.24);

        \node[anchor=west] at (0.57,0.20) {Tx2};
        \node[anchor=east] at (0.48,0.20) {Tx1};
    \end{scope}
    \end{tikzpicture}
    \caption{Effect of the 20 cm adjustment of simulated Tx and Rx locations at $14$~GHz on simulated channel path loss at $14$~GHz.}
    \label{fig:pathloss_correction}
\end{figure}

\begin{table}[t]
    \centering
    \caption{Effects of antenna location adjustments on simulated path gains. Angle of incidence $\theta_{\rm wdw,1}$ and the resulting penetration loss $L_{\rm wdw,1}$ through the triple-glass windows.}
    \begin{tabular}{l l c c}
        \hline\hline
        & Tx/Rx location         & $\theta_{\rm wdw,1}$ [$^\circ$] & $L_{\rm wdw,1}$ [dB]\\
        \multirow{2}{*}{Path~$1$}   & Original & $55.7$& $41.4$\\
                                    & Adjusted & $54.2$& $26.3$\\ \hline
        \multirow{2}{*}{Path~$2$}   & Original & $51.5$& $37.6$\\
                                    & Adjusted & $52.9$& $22.3$\\ \hline
        \multirow{2}{*}{Path~$3$}   & Original & $48.6$& $18.9$\\
                                    & Adjusted & $48.6$& $18.9$\\ \hline
        \multirow{2}{*}{Path~$4$}   & Original & $52.9$& $21.9$\\
                                    & Adjusted & $53.6$& $21.3$\\
        \hline\hline
    \end{tabular}
    \label{tab:adjusted}
\end{table}

\subsection{Correcting for Antenna Location Uncertainty in Channel Sounding}

Figure~\ref{fig:transmission_loss} shows that penetration losses through the modeled layered materials are sensitive to angle of incidence. Combined with uncertainty in Tx and Rx location estimates that were obtained manually during channel sounding, there can be large differences between the simulated and actual penetration losses of a particular path. This can have a large effect on the overall path gain of the simulated channel. 

The effect of this sensitivity on channel gain estimates can be demonstrated by an exemplary link Tx2Rx12 at $14$~GHz band; see Fig.~\ref{fig:ray_tracing_model} for the link geometry. The Rx is in Room~$1$, and the Tx illuminates the building at an angle of incidence approximately at $45^\circ$. Propagation path angles of incidence fall in the strongly oscillating penetration loss region. The PAP of the link is shown in Fig.~\ref{fig:adjusted}. Propagation paths simulated with original Tx and Rx coordinates logged during the measurement campaign and the full floor plan model are shown with green. Alternative paths were obtained by adjusting the Tx and Rx locations within $20 \times 20~{\rm cm^2}$ area so that the resulting path loss of the channel matches the measured best. The effect is summarized in Table~\ref{tab:adjusted}. Trajectories of the $4$ propagation paths are drawn in Fig.~\ref{fig:zoomed_room1}. All of them incident the building at an angle of over $45^\circ$. Path~$1$ is the direct path, while the other are reflections from interior walls of the building. A change of about $1.5^\circ$ in angle of incidence results in $15$~dB change in gains of the path $1$ and $2$. Angular spread of the measured channel is $42.4^\circ$, while the simulated angular spread with original Tx and Rx locations is $53.4^\circ$. With the adjusted Tx and Rx locations taking into account the uncertainty of their estimates during channel sounding, the simulated angular spread becomes $40.9^\circ$.

The same antenna location uncertainty compensation was performed for all simulated Rx locations at $4$ and $14$~GHz bands to improve the path loss estimation accuracy. Results of the $20$~cm adjustment on channel path loss at $14$~GHz band are shown in Fig.~\ref{fig:pathloss_correction}. There is a considerable improvement for links belonging to Tx$2$ because the building is illuminated at an angle of approximately $45^\circ$. Uncertainty in location estimates has a significant effect. The improvement is smaller for the $4$~GHz band because the penetration losses oscillate much less. Overall, RMS errors of simulated path loss decreased from $3.1$ to $2.9$~dB at the $4$~GHz band, and from $5.1$ to $3.6$~dB at the $14$~GHz band. It was necessary to compensate for the antenna location uncertainty in order to perform meaningful comparisons between simulated and measured channels.

\subsection{Channel Simulation Results}
Measured and simulated channel LSPs are presented in Figs.~\ref{fig:LSP4} and \ref{fig:LSP14}. The LSPs are shown for links $1$ through $138$, with links $1$-$69$ corresponding to Tx1 and links $70-138$ to Tx2. Different locations inside the building as seen in Fig.~\ref{fig:ray_tracing_model} are indicated in Figs.~\ref{fig:LSP4}(a) and \ref{fig:LSP14}(a). Shorthand ``R1" stands for Room~$1$, ``R2" for Room~$2$, ``K" for Kitchen and ``C" for Corridor. Shorthand ``RT" stands for ray-tracing.

\subsubsection{Path loss} Measured and simulated path losses are shown in Figs.~\ref{fig:LSP4}(a) and \ref{fig:LSP14}(a). Path losses obtained with both simulated cases follow trends of the measurement well. Overall, the full floor plan simulation replicates measurements better.

Figure~\ref{fig:LSP4}(a) shows that there is a consistent offset of path loss from exteriors RT at $4$~GHz. This can partly be explained with missing reflected paths from walls of R1 and R2 as well as adjacent rooms. Additionally, most of the links are in rooms adjacent to the building exterior. The rooms are empty, and in reality the direct paths and reflections from exterior walls propagating in them are not attenuated. Same consistent but smaller offset be seen for the full floor plan from Tx1 to R1 and R2. It's reasonable to assume that this is a consequence of neglected propagation mechanisms. As seen in Fig.~\ref{fig:ray_tracing_model}, Tx1 transmits to R1 and R2 through many trees. Especially at $4$~GHz they can be expected to contribute paths in the form of scattering and diffraction. There is no consistent offset with full floor plan when Tx2 transmits to R1 and R2. As seen in Fig.~\ref{fig:ray_tracing_model}, there is at most one tree between Tx2 and R2, so there are less sources of errors. Nevertheless, there is good correspondence to measured path losses. 

Large errors from both full floor plan and exteriors RT are observed in the Kitchen at $4$~GHz band. Transmitting from Tx1, the error is an overestimation of path loss suggesting missing propagation paths or mechanisms. A possible explanation is scattering from the trees or the building wall that is almost parallel to the direct path from Tx1 to the Kitchen, neither of which are not included in the simulations. Similarly, diffractions from edges of the double-glass window are not included in the simulations. Nevertheless, the Kitchen is an outlier in terms of errors. Transmitting from Tx2, the exteriors RT has a very large error in the Kitchen. This can be attributed to missing propagation paths that enter the building via R2 and are reflected from its walls toward the Kitchen. These paths are included in the full floor plan RT, hence good agreement with measurements.

Figure~\ref{fig:LSP14}(a) shows good agreement between measurements and full floor plan RT at $14$~GHz. When transmitting from Tx1, exteriors RT path loss exhibits some overestimation particularly in R1 and the Kitchen. This suggests that reflections inside the room and paths from adjacent rooms still contribute to received power, but not as significantly as at $4$~GHz. When transmitting from Tx2, exteriors RT shows significant errors while full floor plan RT reproduces the measured path loss well. This is because the building is now illuminated at an incident angle of approximately $45^\circ$, while for Tx1 it was closer to $0^\circ$. For the exteriors RT case, far fewer paths enter the building, and they seem to fall in high penetration loss parts of Fig.~\ref{fig:transmission_loss}. For the full floor plan RT case, the many reflections from interior walls mean that more paths are likely to fall in the narrow, low penetration loss parts of Fig.~\ref{fig:transmission_loss}, and deliver power to the Rx. The measured path loss is thus significantly lower and better replicated with the full floor plan RT. 

\subsubsection{Delay spread\label{sec:ds}} Measured and simulated delay spreads are shown in Figs.~\ref{fig:LSP4}(b) and \ref{fig:LSP14}(b). The results show that delay spread obtained using full floor plan RT follows its measured counterpart well at both $4$ and $14$~GHz bands. Although the measured trends are replicated well, there are a number of large outlier errors that stand out. The errors can be attributed to missing propagation paths, either due to neglected propagation mechanisms or details of the environment. For example, similarly to path loss, noticeably large errors are seen in the Kitchen for link indices $42-50$. Although not drawn anywhere, the large outliers there are explained by a RT propagation path that takes a very long trajectory via R1 and Corridor to the Kitchen. Delay spread obtained with exteriors RT consistently underestimates the measured ones, although it seems to somewhat follow the trend. This can be attributed to the large number of missing propagation paths from the building interior. Reflections from only the exterior walls are not enough to accurately replicate the measured delay spread, but they result in delay spreads of slighly lower levels.

\subsubsection{Angular spread\label{sec:as}} Measured and simulated angular spreads are shown in Figs.~\ref{fig:LSP4}(c) and \ref{fig:LSP14}(c). Full floor plan RT replicates the measured angular spread of many individual links and trends well. Exteriors RT fails to reproduce the measured angular spread completely. The reason for exteriors RT having some success with delay spread but none at all with angular spread is that in delay domain, paths from the exteriors are realistically spread apart. In the angular domain, they aren't at all because missing interior means to disregard all propagation paths coming from azimuth angular range between $180^\circ$ to $360^\circ$ shown on Fig.~\ref{fig:ray_tracing_model}. The angular spread is therefore significantly underestimated. 

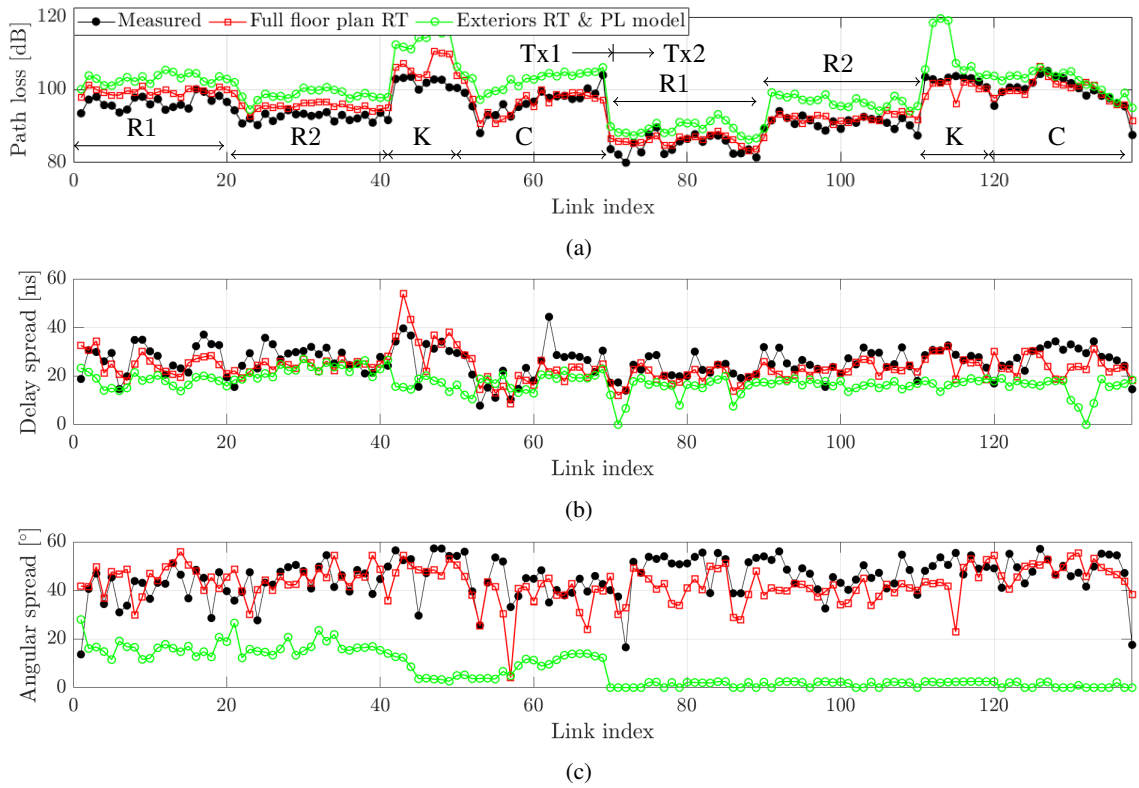
\begin{figure*}
\begin{subfigure}[b]{\linewidth}
    \begin{tikzpicture}
    \node[anchor=south west,inner sep=0] (image) at (0,0) {\includegraphics[width=\linewidth]{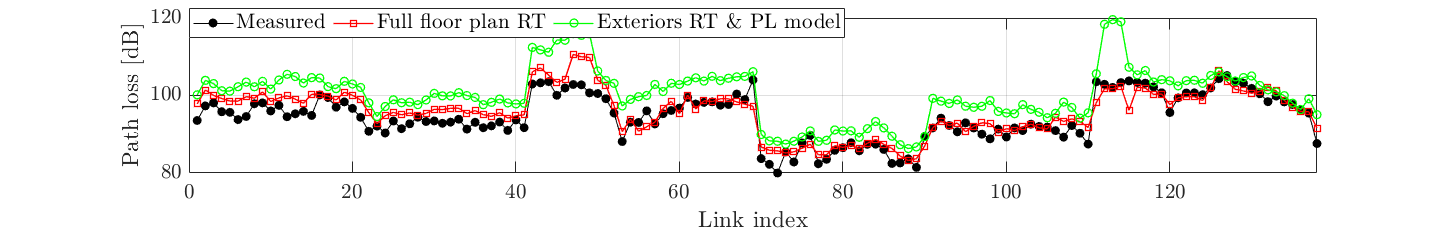}};
    \begin{scope}[x={(image.south east)},y={(image.north west)}]

        \draw [<->] (0.13,0.34) -- (0.24,0.34);
        \draw [<->] (0.245,0.3) -- (0.36,0.3);
        \draw [<->] (0.36,0.3) -- (0.41,0.3);
        \draw [<->] (0.41,0.3) -- (0.52,0.3);
        
        \node[anchor=south] at (0.18,0.34) {R$1$};
        \node[anchor=south] at (0.3,0.3) {R$2$};
        \node[anchor=south] at (0.385,0.3) {K};
        \node[anchor=south] at (0.46,0.3) {C};

        \draw [<->] (0.525,0.54) -- (0.63,0.54);
        \draw [<->] (0.635,0.63) -- (0.75,0.63);
        \draw [<->] (0.75,0.3) -- (0.8,0.3);
        \draw [<->] (0.8,0.3) -- (0.9,0.3);
        
        \node[anchor=south] at (0.57,0.54) {R$1$};
        \node[anchor=south] at (0.69,0.63) {R$2$};
        \node[anchor=south] at (0.775,0.3) {K};
        \node[anchor=south] at (0.85,0.3) {C};

        \draw [->] (0.525,0.76) -- (0.555,0.76);
        \draw [<-] (0.525,0.76) -- (0.495,0.76);
        \draw [-] (0.525,0.72) -- (0.525,0.8);

        \node[anchor=west] at (0.555,0.76) {Tx2};
        \node[anchor=east] at (0.495,0.76) {Tx1};
    \end{scope}
    \end{tikzpicture}
    \caption{}
\end{subfigure}
\begin{subfigure}[b]{\linewidth}
    \begin{tikzpicture}
    \node[anchor=south west,inner sep=0] (image) at (0,0) {\includegraphics[width=\linewidth]{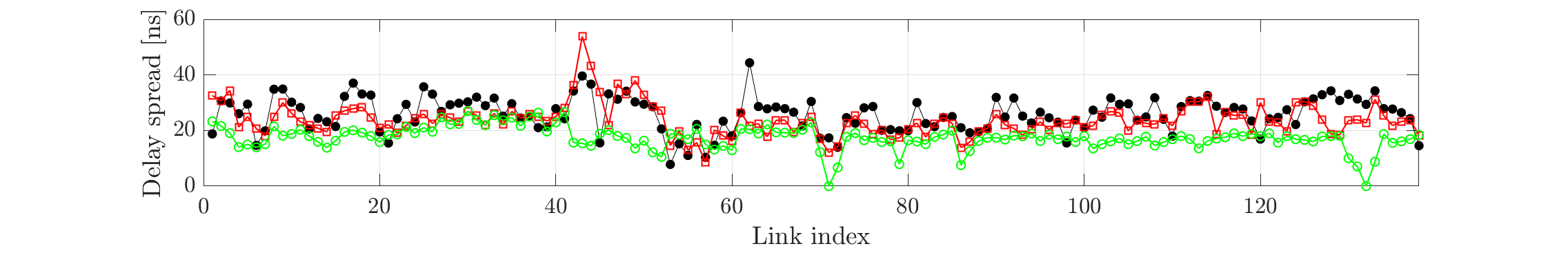}};
    \begin{scope}[x={(image.south east)},y={(image.north west)}]

    \end{scope}
    \end{tikzpicture}
    \caption{}
\end{subfigure}
\begin{subfigure}[b]{\linewidth}
    \begin{tikzpicture}
    \node[anchor=south west,inner sep=0] (image) at (0,0) {\includegraphics[width=\linewidth]{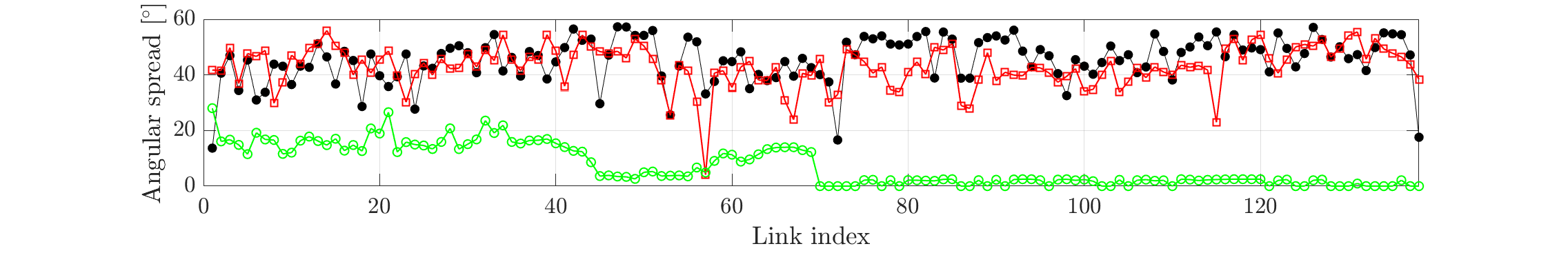}};
    \begin{scope}[x={(image.south east)},y={(image.north west)}]

    \end{scope}
    \end{tikzpicture}
    \caption{}
\end{subfigure}
\caption{Measured and simulated $4$~GHz channel path losses (a), delay spreads (b) and angular spreads (c).}\label{fig:LSP4}
\end{figure*}
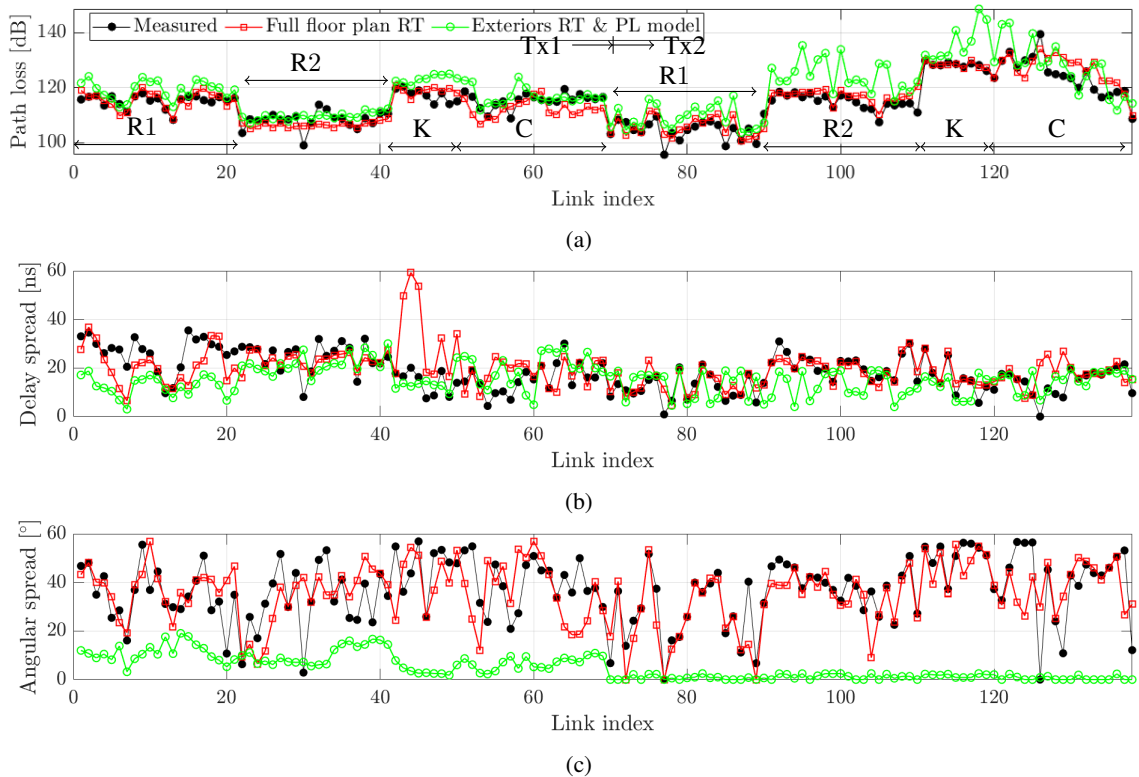
\begin{figure*}
\begin{subfigure}[b]{\linewidth}
    \begin{tikzpicture}
    \node[anchor=south west,inner sep=0] (image) at (0,0) {\includegraphics[width=\linewidth]{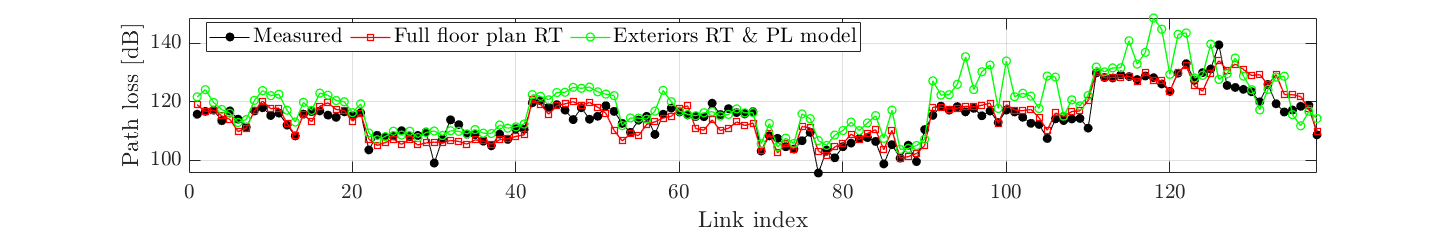}};
    \begin{scope}[x={(image.south east)},y={(image.north west)}]

        \draw [<->] (0.13,0.31) -- (0.25,0.31);
        \draw [<->] (0.255,0.6) -- (0.36,0.6);
        \draw [<->] (0.36,0.3) -- (0.41,0.3);
        \draw [<->] (0.41,0.3) -- (0.52,0.3);
        
        \node[anchor=south] at (0.18,0.31) {R$1$};
        \node[anchor=south] at (0.3,0.6) {R$2$};
        \node[anchor=south] at (0.385,0.3) {K};
        \node[anchor=south] at (0.46,0.3) {C};

        \draw [<->] (0.525,0.55) -- (0.63,0.55);
        \draw [<->] (0.635,0.3) -- (0.75,0.3);
        \draw [<->] (0.75,0.3) -- (0.8,0.3);
        \draw [<->] (0.8,0.3) -- (0.9,0.3);
        
        \node[anchor=south] at (0.57,0.54) {R$1$};
        \node[anchor=south] at (0.69,0.3) {R$2$};
        \node[anchor=south] at (0.775,0.3) {K};
        \node[anchor=south] at (0.85,0.3) {C};

        \draw [->] (0.525,0.76) -- (0.555,0.76);
        \draw [<-] (0.525,0.76) -- (0.495,0.76);
        \draw [-] (0.525,0.72) -- (0.525,0.8);
        
        \node[anchor=west] at (0.555,0.76) {Tx2};
        \node[anchor=east] at (0.495,0.76) {Tx1};
    \end{scope}
    \end{tikzpicture}
    \caption{}
\end{subfigure}
\begin{subfigure}[b]{\linewidth}
    \begin{tikzpicture}
    \node[anchor=south west,inner sep=0] (image) at (0,0) {\includegraphics[width=\linewidth]{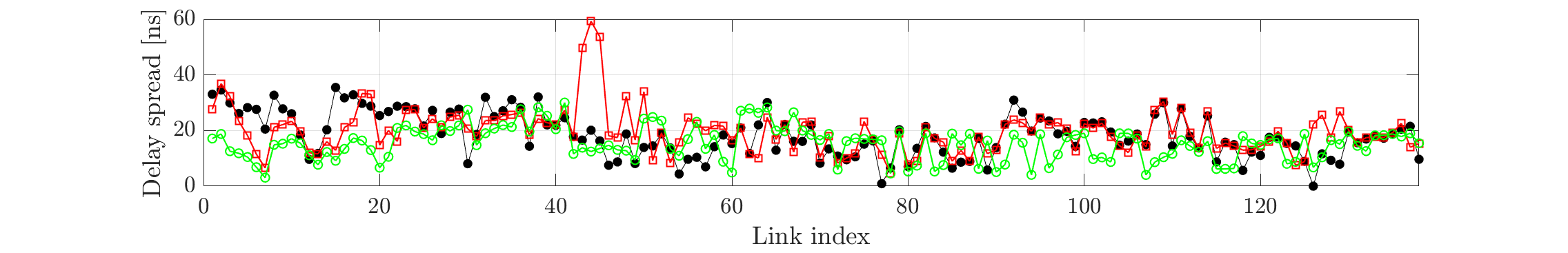}};
    \begin{scope}[x={(image.south east)},y={(image.north west)}]

    \end{scope}
    \end{tikzpicture}
    \caption{}
\end{subfigure}
\begin{subfigure}[b]{\linewidth}
    \begin{tikzpicture}
    \node[anchor=south west,inner sep=0] (image) at (0,0) {\includegraphics[width=\linewidth]{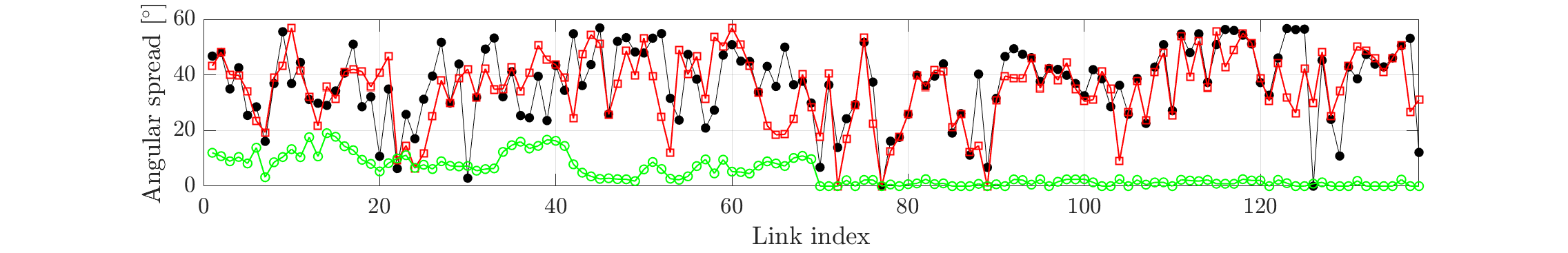}};
    \begin{scope}[x={(image.south east)},y={(image.north west)}]

    \end{scope}
    \end{tikzpicture}
    \caption{}
\end{subfigure}
\caption{Measured and simulated $14$~GHz channel path losses (a), delay spreads (b) and angular spreads (c).}\label{fig:LSP14}
\end{figure*}

\subsubsection{Paths Originating From Exterior and Interior Walls}

\begin{figure*}[t]
    \begin{tikzpicture}
    \node[anchor=south west,inner sep=0] (image) at (0,0) {\includegraphics[width=0.5\linewidth]{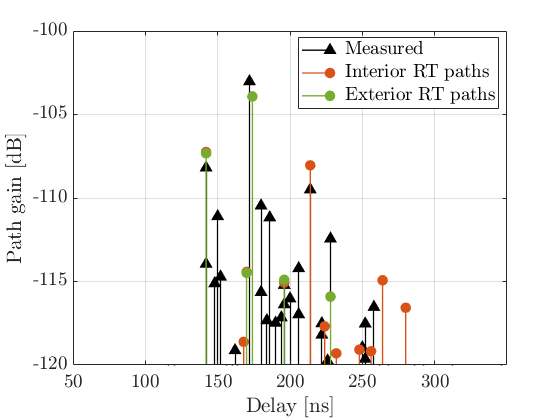}};
    \begin{scope}[x={(image.south east)},y={(image.north west)}]
    
        \draw [thick,->] (0.24,0.70) --  (0.35,0.60);
        \node[anchor=south] at (0.24,0.70) {Direct path};
        
        \draw [thick,->] (0.34,0.85) --  (0.42,0.8);
        \node[anchor=south] at (0.34,0.85) {Parking structure};
        
    \end{scope}
    \end{tikzpicture}
    \begin{tikzpicture}
    \node[anchor=south west,inner sep=0] (image) at (0,0) {\includegraphics[width=0.5\linewidth]{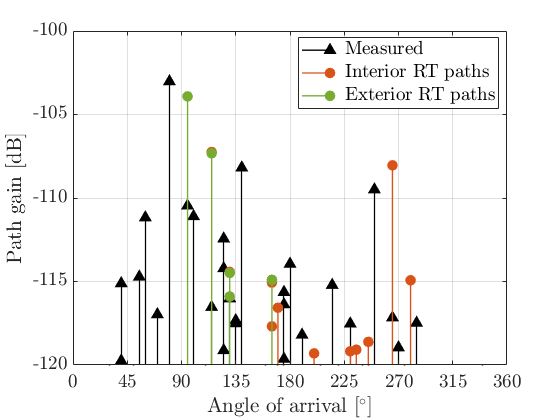}};
    \begin{scope}[x={(image.south east)},y={(image.north west)}]
    
        \draw [thick,->] (0.5,0.68) --  (0.40,0.65);
        \draw [thick,->] (0.5,0.68) --  (0.44,0.62);
        \node[anchor=south] at (0.5,0.67) {Direct path};
        
        \draw [thick,->] (0.4,0.85) --  (0.33,0.8);
        \node[anchor=south] at (0.37,0.85) {Parking structure};
        
    \end{scope}
    \end{tikzpicture}
    \caption{Measured and simulated power delay profile (a) and power angular profile (b) of Tx1Rx3 at $4$~GHz band. Simulated paths originating from exterior and interior walls of the site are indicated separately.}
    \label{fig:PDPPAP}
\end{figure*}

Having observed that simulations using only building exteriors underestimate delay and angular spreads, measured and full floor plan RT simulated propagation paths of a specific link are studied next. Figure~\ref{fig:PDPPAP} shows the PDP and PAP of link Tx1Rx3 at $4$~GHz band. The Rx3 antenna is located in R1 at the corner of the office building, while the Tx1 antenna on the other side of a cluster of trees. The Rx3 location inside R1 is shown in Fig.~\ref{fig:zoomed_room1}, with Tx1 being approximately toward $135^\circ$. The strongest paths shown with green circles originate from the exterior walls. Figure~\ref{fig:PDPPAP}(a) shows that after approximately $200$~ns, paths bounded on interior walls are required to approximate the measured PDP. Figure~\ref{fig:PDPPAP}(b) shows that paths originating from the exterior walls arrive only from angles between $0^\circ$ and $180^\circ$. To reproduce the PAP from approximately $180^\circ$ to $360^\circ$, paths reflected from interior walls are required. 

Measured delay and angular spreads of the link Tx1Rx3 are $29.9$~ns and $46.9^\circ$. With the full floor plan, the simulated delay and angular spreads are $34.6$~ns and $49.8^\circ$. While with the exterior wall only, the simulated delay and angular spreads are $19.4$~ns and $17.0^\circ$. This further demonstrates that interior paths are required for increased accuracy in reproducing the delay and angular spread values.

\begin{table*}[t]
    \centering
    \caption{Measured and simulated LSPs at $4$ and $14$~GHz bands. The Table reports mean value of the LSP, its standard deviation, and absolute and relative mean and RMS errors in raw numbers and percentage values; the latter is shown in parentheses.}
    \begin{tabular}{l l c c c c c c}
\hline\hline
        Frequency band                           &                   & \multicolumn{3}{c}{$4$~GHz} & \multicolumn{3}{c}{$14$~GHz}\\ \hline
                                                    &                   & Measured & Full floor plan RT & Exteriors RT & Measured & Full floor plan RT & Exteriors RT\\ \hline
        \multirow{4}{*}{PL [dB]}                    & Mean value        & $94.6$  & $95.9$          & $100.3$  & $114.7$ & $115.0$ & $119.2$\\
                                                    & Standard deviation& $5.9$   & $5.8$           & $6.8$   & $7.8$   & $8.2$   & $9.4$\\ 
                                                    & Mean error        & n/a   & $1.4$           & $5.8$   & n/a   & $0.3$   & $4.5$\\
                                                    & RMS error         & n/a   & $2.9$           & $6.6$   & n/a   & $3.6$   & $7.0$\\ \hline
        \multirow{4}{*}{$\tau_{\rm RMS}$ [ns]}      & Mean value        & $26.0$  & $23.8$          & $17.4$  & $18.5$  & $19.8$  & $15.4$\\
                                                    & Standard deviation& $6.2$   & $5.8$           & $4.2$   & $7.9$   & $8.1$   & $6.1$\\ 
                                                    & Mean error        & n/a   & $-2.1$ ($-8.3\%$) & $-8.6$ ($-33.1\%$)& n/a   & $1.2$ ($6.6\%$)   & $-3.1$ ($-17.0\%$) \\
                                                    & RMS error         & n/a   & $6.1$ ($23.6\%$)  & $11.0$ ($42.3\%$) & n/a   & $8.5$ ($45.8\%$) & $9.5$ ($51.1\%$)\\ \hline
        \multirow{4}{*}{$\phi_{\rm RMS}$ [$^\circ$]}& Mean value        & $45.4$  & $43.0$          & $7.3$   & $36.9$  & $35.5$  & $4.9$\\
                                                    & Standard deviation& $8.2$   & $7.5$           & $7.2$   & $13.4$  & $12.7$  & $5.0$\\ 
                                                    & Mean error        & n/a   & $-2.4$ ($-5.4\%$) & $-38.2$ ($-84.0\%$)& n/a  & $-1.4$ ($-3.8\%$) & $-32.0$ ($-86.7\%$)\\
                                                    & RMS error         & n/a   & $9.7$ ($21.4\%$) & $40.1$ ($88.3\%$) & n/a   & $12.0$ ($32.6\%$) & $35.1$ ($95.0\%$) \\ 
                                                    \hline\hline
                                                    
    \end{tabular}
    \label{tab:LSPs}
\end{table*}

\begin{table*}[t]
    \centering
    \caption{Comparison of this work to previously published O2I simulations that were validated with measurements. The Table reports mean error (ME) and RMS error (RMSE) for three LSPs. Entry of n/a means no value was reported.}\label{tab:prev_works}
    \begin{tabular}{l l l l l l l l l}
    \hline\hline
        \multirow{2}{*}{Reference}                               & \multirow{2}{*}{Frequency} & \multicolumn{2}{c}{Path loss [dB]} & \multicolumn{2}{c}{Delay spread [ns (Rel.)]} & \multicolumn{2}{c}{Angular spread [$^\circ$ (Rel.)]} & \multirow{2}{*}{Note}\\
        & & ME & RMSE & ME & RMSE & ME & RMSE \\ \hline
        \multirow{2}{*}{\cite{Roche10_EURASIP}} & $3.5$~GHz & $0.09$ & $2.39$ & \multicolumn{2}{c}{\multirow{2}{*}{n/a}} & \multicolumn{2}{c}{\multirow{2}{*}{n/a}} & RT outside \& finite difference\\
                                                & $2.4$~GHz & $0.21$ & $1.17$ & & & & & inside with one level floor plan.\\ \hline
        \multirow{1}{*}{\cite{Wang00_TAP}}      & $2.4$~GHz & n/a    & $1.31$ & \multicolumn{2}{c}{n/a} & \multicolumn{2}{c}{n/a} & \multirow{1}{*}{RT to building w/ single room.}\\ \hline
        \multirow{2}{*}{\cite{Degli-Esposti17_Access}} & $858$~MHz   & n/a& $2.74$ & \multicolumn{2}{c}{\multirow{2}{*}{n/a}} & \multicolumn{2}{c}{\multirow{2}{*}{n/a}} & No floor plan, PL model\\
                                                       & $1.935$~GHz & n/a& $3.95$ & & & & & applied inside building.\\ \hline
        \multirow{1}{*}{\cite{DegliEsposti17_AWPL}} & $850$~MHz   & $-2.1$ & n/a& \multicolumn{2}{c}{\multirow{1}{*}{n/a}} & \multicolumn{2}{c}{\multirow{1}{*}{n/a}} & \multirow{1}{*}{Multi-story virtual floor plan.}\\ \hline
        \multirow{2}{*}{\cite{Jimenez_CONCAPAN17}} & $0.85$~GHz   & $0$-$20$& n/a & \multicolumn{2}{c}{\multirow{2}{*}{n/a}} & \multicolumn{2}{c}{\multirow{2}{*}{n/a}} & Floor plans, mean error\\
                                                       & $1.9$~GHz & $3$-$14$& n/a & & & & & ranges across $5$ floors.\\ \hline
        \multirow{2}{*}{This work} & $4$~GHz & $1.4$ & $2.9$ & $-2.1$ $(-8.3\%)$ & $6.1$ $(23.6\%)$ & $-2.4$ $(-5.4\%)$ & $9.7$ $(21.4\%)$ & Point cloud RT using full floor\\
                                   & $14$~GHz & $0.3$ & $3.6$& $1.2$ $(6.6\%)$ & $8.5$ $(45.8\%)$ & $-1.4$ $(-3.8\%)$ & $12.0$ $(32.6\%)$ & plan of one building level.\\
    \hline\hline
    \end{tabular}
    \label{tab:my_label}
\end{table*}

\subsubsection{Estimation Errors of LSPs}
Comparison of the measured and simulated LSPs is summarized in Table~\ref{tab:LSPs}. The mean and standard deviations of the measured and simulated LSPs are shown, as well as the mean and RMS errors of the simulated results against measurements derived from all Tx-Rx links. The positive mean error means that the simulated LSPs are greater than those of measurements. The mean error encompasses accuracy of the channel simulation overall, whereas the RMS error indicates link-specific accuracy. Standard deviation of the LSP provides the range of values in our O2I site.

For path loss a mean error of $1.4$ and $0.3$~dB is achieved using the full floor plan RT at the two bands. Mean error is less than $6$~dB for both frequency bands using exteriors RT. Similarly, the RMS errors of path loss are much lower for the full floor plan RT over exteriors RT. Using full floor plan RT RMS errors of $2.9$ and $3.6$~dB are achieved, while for exteriors RT they are approximately doubled. While the exteriors RT errors in path loss are clearly higher, it can still be said that path loss inside the building can be reproduced to a reasonable degree.

For delay and angular spreads describing multipath richness of the environment, full floor plan RT achieves a relative mean error of under $10\%$ at both frequency bands. For reasons discussed in Sections~\ref{sec:ds} and \ref{sec:as} the mean errors are much higher for exteriors RT. The relative RMS error of delay and angular spreads are high at both frequency bands, tens of percentage points, even when using the full floor plan. This suggests that while the channel is well reproduced on average, the link-specific values are much more difficult to duplicate. This can be explained by simplifications made during RT simulations. Wooden window frames and a few concrete supporting structures in the building facade were ignored by assuming that propagation paths always enter through a window. For example, a reflection from the parking structure, shown in Fig.~\ref{fig:ray_tracing_model}, could be blocked and heavily attenuated by a concrete pillar for one Tx-Rx link but not the next one in the measurements. This effect cannot be reproduced in the simulation due to the assumed homogenized window wall of the building exterior. It is assumed that tree canopies are homogeneous. The mean effect is well reproduced, but in reality a large branch can block a propagation path while another one passes through some leaves. Similarly it was assumed that the interior walls are homogeneous double-plasterboard walls with an air gap. In reality there are some variations in materials, and the air gaps may contain electrical installations and supporting structures. Another simplification of the RT simulations was calculation of only direct and reflected paths, which is a reasonable explanation for the high RMS error of delay and angular spreads. They are far more sensitive to individual propagation paths and their gains. Nevertheless, even with the simplification, the radio channels are well reproduced, suggesting that the direct and reflected paths are still clearly dominant propagation mechanisms over diffuse scattering and diffractions. While diffraction is known to be an important mechanism in non-line-of-sight conditions, many of the Rx locations are essentially within obstructed line-of-sight of the Tx with a window and some canopy between them. However, no more can be said without further study.

Finally, the measured standard deviation of LSPs are reproduced well by full floor plan RT. They vary in a range that is similar to the measured results. Using exteriors RT, the standard deviation is not as well reproduced. This can be explained with the less realistic modeling of multipath richness.

\subsection{Comparisons to Previous Works}
Results presented in this work are summarized in Table~\ref{tab:prev_works} and compared to previously published O2I simulations that were validated against measurements. While a large number of O2I simulations have been performed and published, validation against measurements is lacking in general. Moreover, this work for the first time in the literature performs the validation for above-$6$~GHz radio frequency. The Table shows that path loss RMS errors achieved in this work are in line with earlier works. The table also includes comparisons of the delay and angular spreads for the first time in the literature, showing sub-$10\%$ mean errors of the simulated values in reference to the measurements. It is difficult to assess how good or bad the RMS errors of tens of \%-points are due to lack of earlier publications. The absolute values are for the most part less than approximately $10$~ns or $10^\circ$.

\section{Conclusion}\label{sec:conclusion}

This work presents comparisons of O2I ray-tracing simulations and measurements performed at $4$ and $14$~GHz bands utilizing a laser-scanned point cloud. The measurements were conducted at a typical office building with windows covering the exterior and many rooms separated with plasterboard walls. Measured and simulated direct path excess losses were first used to determine distance-dependent propagation losses inside tree canopies and the office building interior. The values were then applied to ray-traced propagation paths with up to $4$ reflections. Ray-tracing simulations were performed with two set-ups; using \textit{only exterior walls} with no knowledge of building interior and using \textit{full floor plan} of the level housing Rx antennas. For the former case, the determined interior loss was applied to propagation inside the building in place of a specific floor plan. 

Results from the two ray-tracing set-ups were compared with their measured counterparts in terms of channel LSPs to benchmark the accuracy and importance of building floor plan in reproducing measured channels. The results confirm that it is possible to reproduce the measured path loss to a reasonable degree with a mean error of less than $6$~dB at both $4$ and $14$~GHz bands. A full floor plan reduces the error to under $1.5$~dB at both frequency bands. The path loss errors achieved in this work are in line with earlier publications utilizing various approaches to O2I simulations, although the validation at above-$6$~GHz band is a new achievement.

Conversely, delay and angular characteristics of the channel cannot be accurately reproduced without a floor plan of the building. To this effect, this work reports a first validation of simulated delay and angular spreads at both above and below-$6$~GHz bands. A relative mean error of less than $10\%$ can be achieved after a careful consideration of the building window types. It was shown that penetration losses of multi-layered insulating windows fluctuate strongly across incident angles. This in turn results in large changes in simulated propagation path gains and coverage due to high angular selectivity of the window. This was compensated with small adjustments of the simulated Rx antenna locations to accurately replicate measurements. The RMS error of simulated delay and angular spreads was much larger than mean errors at both bands, but maintained under approximately $10$~ns and $10^\circ$. This indicates that while good overall results were achieved with interactions limited to reflections, further study of diffraction and diffuse scattering as O2I propagation mechanisms is required.


In the future energy efficiency requirements imposed on buildings can be expected to increase further. This can lead to, e.g., in cold climate countries such as Finland, an increased number of triple-glass windows, upgrades to four glass-panes, and possibly multiple metal films for added insulation and penetration loss. This warrants special considerations and further study of planning and simulating O2I coverage.


\section*{Acknowledgement}
The research leading to these results received funding from the LuxTurrim5G project funded by the participating companies and Business Finland. The results presented in this paper has been partly supported by the Academy of Finland project "Signal-Transmissive-Walls with Embedded Passive Antennas for Radio-Connected Low-Energy Urban Buildings (STARCLUB)", \#323896. The authors would like to thank the Nokia Bell-Labs for providing us with the opportunity to perform outdoor-to-indoor channel sounding in the Nokia campus in Karaportti, Espoo, Finland, and Dr.\ Sinh Nguyen, Mr.\ Usman Virk, Dr.\ Vasilii Semkin, Mr.\ Jyri Putkonen, Mr.\ Yuan Liu, Ms.\ Le Hao and Dr.\ Pekka Wainio for help during the measurements and data post-processing.

\bibliographystyle{IEEEtran}
\bibliography{refsnew,refKatsuold}
\end{document}